\documentclass[a4paper,twocolumn,showkeys]{revtex4}  %preprint,showkeys, onecolumn, prl,showpacs,twocolumn,13pt,
\usepackage{latexsym}
\usepackage[colorlinks=true,linkcolor=blue,anchorcolor=blue,citecolor=blue]{hyperref}
\usepackage[hmargin={1cm,1cm}, vmargin={2cm,2cm}]{geometry}
\usepackage{graphicx}
\usepackage{graphicx,amsfonts,amsmath,amssymb,amsthm,amscd,ragged2e}
\usepackage{bm}
\usepackage{xr}
\usepackage{indentfirst}
\usepackage{appendix}
\usepackage{changepage}
\usepackage{multirow}
\usepackage{appendix}
\renewcommand{\thefigure}{\arabic{figure}}

\date{\today}

\begin{document}

%\title{\bf Stability factors of motor-clutch model on compliant substrates under load}

\title{\bf Factors influencing the stability of the motor-clutch model on compliant substrates under external load}

\author{Beibei Shen and Yunxin Zhang} \email[Email: ]{xyz@fudan.edu.cn}
\affiliation{Shanghai Key Laboratory for Contemporary Applied Mathematics, School of Mathematical Sciences, Fudan University, Shanghai 200433, China.}

\begin{abstract}
Cellular migration is crucial for biological processes including embryonic development, immune response, and wound healing.
The myosin-clutch model is a framework that describes how cells control migration through the interactions between myosin, the clutch mechanism, and the substrate. 
This model is related to how cells regulate adhesion, generate traction forces, and move on compliant substrates. 
In this study, we present a five-dimensional nonlinear autonomous system to investigate the influences of myosin, clutches, substrate, and external load on the system's stability. 
Moreover, we analyze the effects of various parameters on fixed points and explore the frequency and amplitude of the limit cycle associated with oscillations.
We discovered that the system demonstrates oscillatory behavior when the velocity of the myosin motor is relatively low, or when the ratio of the motor attachment rate to motor detachment rate is relatively high. 
The external load shares a fraction of the force exerted by myosin motors, thereby diminishing the force endured by the clutches.
Within a specific range, an increase in external load not only diminishes and eventually eliminates the region lacking fixed points but also decelerates clutch detachment, enhancing clutch protein adherence.
\end{abstract}


%\pacs{111,eee}

\maketitle

\section{Introduction}
Cellular migration plays a pivotal role in a wide array of biological processes, including embryonic development, the immune response, and wound healing \cite{petrie2009random,de2005integrin,paszek2005tensional,yamaguchi2005cell,paluch2016focal,plotnikov2013guiding}. 
This process is intricately regulated by the cell's ability to sense and respond to the mechanical stiffness of its surroundings.     
Through the use of their contractile and adhesive molecular machinery, cells apply and transmit forces across the extracellular matrix, converting these mechanical interactions into biochemical signals that influence cell behavior and fate \cite{moore2010stretchy,chan2008traction,elosegui2014rigidity,case2015integration}.
Key to this regulatory mechanism is the actomyosin complex, comprised of actin filaments, myosin motors, and adaptor proteins.    
These adaptor proteins form crucial links to transmembrane proteins, thereby establishing connections with the cell's microenvironment and facilitating migration \cite{elosegui2018control,blanchoin2014actin,charras2014physical}.
The nuanced interplay between mechanical sensing and the molecular components driving cell migration underscores the complexity of cellular responses in various physiological contexts, with the mechanical stiffness of the environment playing a significant, albeit variable, role in dictating cell behavior and destiny \cite{lo2000cell,califano2010substrate}.

The molecular assembly facilitating cell migration, termed the adhesion complex, exhibits significant dynamism, primarily through the kinetics of attachment and detachment among components linking the cytoskeleton to the extracellular matrix.   This dynamic behavior is crucial for the orchestrated movement of cells.
The combined actions of myosin motors generating contractile forces and the polymerization of actin filaments against the cell membrane induce a \lq retrograde flow' of actin towards the cell's center.
Central to this process is the \lq molecular clutch' hypothesis, which suggests that focal adhesions (FAs) act as mechanical clutches.      These clutches serve as dynamic linkages, facilitating the transmission of forces from the actin filaments to the transmembrane proteins and effectively converting the retrograde actin flow into forward cellular motion \cite{elosegui2016mechanical}. Furthermore, cells use this molecular clutch mechanism to sense and respond to environmental stiffness by transmitting forces to the extracellular matrix \cite{elosegui2014rigidity,plotnikov2012force,gupta2015adaptive,oakes2014geometry} and transforming these mechanical interactions into biochemical signals \cite{dupont2011role}.  

Recent experiments using time-lapse traction force microscopy have revealed that the local forces applied by individual focal adhesions (FAs) vary over time and space, suggesting that they periodically tug on the extracellular matrix/substrate \cite{plotnikov2013guiding,plotnikov2012force}.  
Mature focal adhesions exhibit both stable and dynamic states, with the stable state characterized by invariant traction across both space and time, and the dynamic state involving fluctuating traction forces that suggest a tugging mechanism on the extracellular membrane. Such  force fluctuations might serve as a molecular mechanism through which cells can precisely regulate their movement in reaction to environmental signals \cite{wu2017two,gardel2010mechanical}. However, integrating these fluctuations into a comprehensive model of cell migration remains a challenge, as the underlying mechanisms are not fully understood. Myosin contractility plays a crucial role in generating these fluctuations, with existing models suggesting that the combined activity of myosin motors and elastic substrates can lead to spontaneous oscillatory behaviors in local contractile units \cite{grill2005theory}.

 Previous theoretical models predicted the spontaneous, directed movements of actin motor proteins \cite{julicher1995cooperative} and subsequently a stick-slip type of dynamics within the motor-clutch paradigm \cite{chan2008traction,bangasser2013master}.
A stochastic model of cell force transmission based on the motor-clutch hypothesis, presented by Chan and Odde \cite{chan2008traction}, considers the force-velocity relationship of the myosin motors and incorporates the load-and-fail dynamics of cellular adhesions.
However, these models either assumed that the forces exerted by the stress fibers on FAs are constant \cite{sabass2010modeling,chan2008traction}, or they did not take into account the role of myosin contractility and the attachment-detachment dynamics \cite{bangasser2013determinants}.
Despite the fact that the various components of cell migration process are well-known, the measurement of mechanical forces exhibits substantial variability at the cellular level \cite{kurzawa2017dissipation,meili2010myosin,rape2011regulation,plotnikov2012force}, making it imperative to decipher the crucial parameters that govern these dynamics. 

Our model, by considering a range of biologically relevant parameters, exhibits a rich array of dynamic states. The structural framework for describing our model was derived from the work of D. Ghosh et al. \cite{ghosh2022deconstructing}.
In contrast to the work of D. Ghosh et al., which treats the motor-clutch and substrate sectors as separate blocks, we argue that these sectors form an integral block. Their dynamics, we believe, should be analyzed in conjunction. 
Additionally, research has demonstrated that the motor-clutch system inherently senses and responds to the mechanical stiffness of its local environment \cite{yeung2005effects,engler2006matrix,chan2008traction}. For a deeper understanding of stiffness sensing, it is crucial to explore how substrate stiffness impacts the system within the \lq motor-clutch' force transmission mechanism.

In this study, we introduce a set of coupled ordinary differential equations that serve as a system model for describing and simulating the dynamics of intracellular actomyosin activity and the complex biomechanical interactions between the cell and its external environment.
We investigate the system's dynamical stability by considering various factors, including myosin motors, molecular clutches, compliant substrates, and external loads.
A variation of the relevant parameters of each factor gives rise to a multitude of dynamical states.
Specifically, from the perspective of myosin motors, we pay particular attention to how myosin II activity affects dynamical stability.
The myosin II activity is characterized by the velocity of the attached myosin motors, as well as by its dynamics of attachment and detachment with the actin filament. 
Through the proposed system, we validate the significant influence of myosin activity on oscillatory behaviors.
Furthermore, we investigate the impact of parameter variations on fixed points, and we delve into local fluctuations induced by spontaneous oscillations through a calculation of the frequency and amplitude of the limit cycle.
Through in-depth analysis, our model not only complements the existing theoretical framework but also provides a new perspective for understanding cellular mechanical behavior.

\begin{figure}[htbp]
	\includegraphics[scale=0.108]{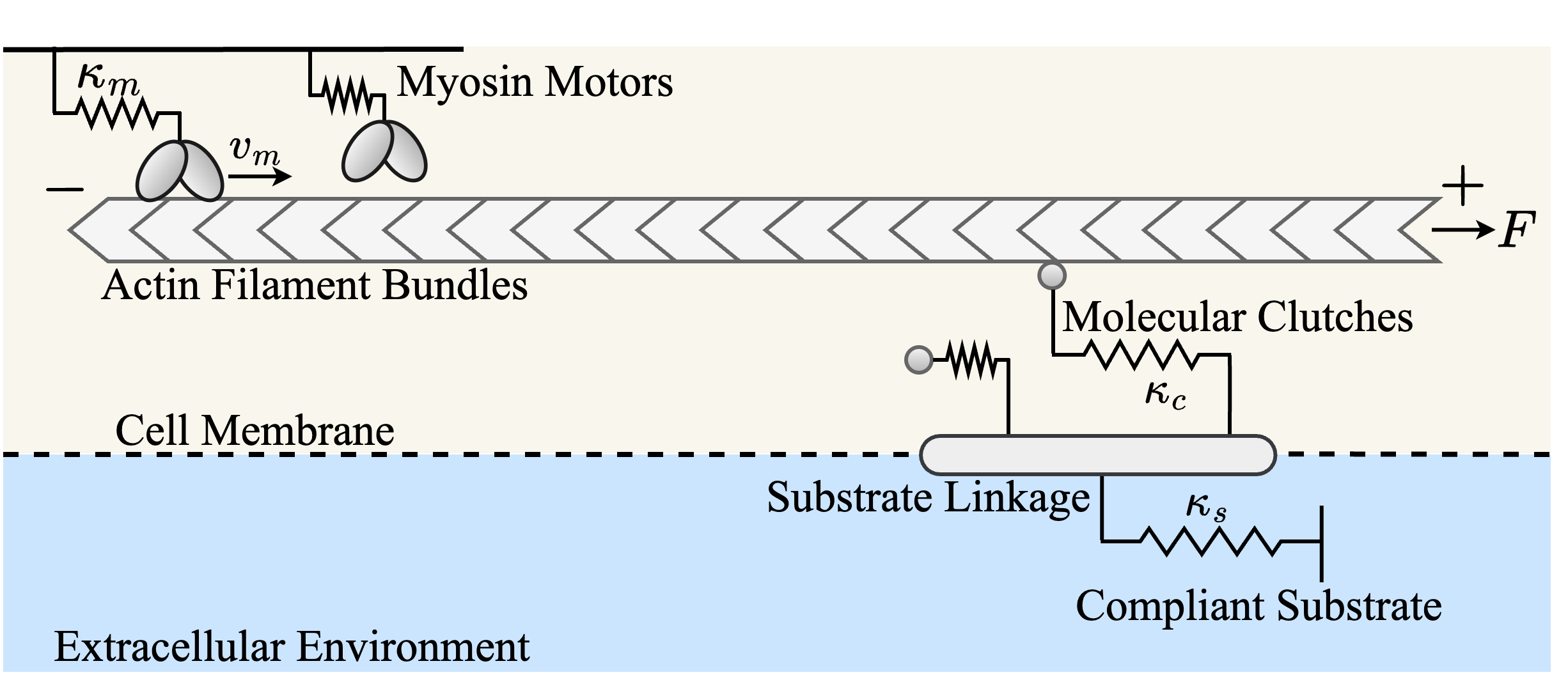}\\
	\caption{\label{myosin_clutch} A model for motor-clutch motility on compliant substrates under external load $F$. Myosin motors exert forces on inextensible actin filament bundles, inducing their movement in the retrograde direction. Molecular clutches reversibly engage the actin filament bundles to resist retrograde flow and transmit forces to a compliant substrate in the extracellular environment.
		Myosin motors, molecular clutches, and compliant substrates are modeled as linear elastic springs, each characterized by its mechanical stiffness: $\kappa_{m}$ for the motors, $\kappa_{c}$ for the clutches, and $\kappa_{s}$ for the substrate. The symbols $+$ and $-$ are used to indicate the anterograde and retrograde directions, respectively.}
\end{figure}
\section{Methods}
\subsection{Model Description}

In this paper, we explore a model of myosin-clutch motility on compliant substrates under external load $F$, treating the motor-clutch mechanism and the substrate as a unified entity for our investigation. Considering an average availability of $N_m$ myosin motors, with $n_m$ of these motors being attached to the actin bundle.
They are modeled as stretchable linear springs, and the extension of the $i$-th myosin motor, denoted as $y^i$, is directly proportional to the load force exerted, represented as $f_l^i=\kappa_m y^i$. Here, $\kappa_m$ is referred to as the spring stiffness.
Consequently, the average load force exerted by the myosin motors can be expressed as
$f_l=\kappa_m y$, where $y$ represents the mean extension of these attached motors, as $y=\frac{1}{n_m} \sum_{i=1}^{n_m} y^i$.

Myosin motors undergo attachment-detachment dynamics to and from the actin filament bundles, a process driven by the energy released during ATP hydrolysis.
The contractility and the resultant force generation of myosin motors are dependent on the dynamics of attachment and detachment to actin filaments \cite{stam2015isoforms,koenderink2018architecture}. One end of the spring is anchored, whereas the other end attaches to and detaches from the F-actin bundle with rates $\omega_{a}$
and $\omega_{d}$, respectively. 
Similar to \cite{Muller4609,ZhangFisher2010}, the detachment rate
$\omega_{d}$ of motors from the actin filament is assumed to be of the form
$
\omega_d=\omega_d^0 \exp \left(\left|f_l\right| / f_d\right)
$
where parameter  $f_d$ represents the detachment force, $\left | f_{l} \right | $ is the load force, and $\omega_d^0$ is the bare detachment rate.
Then, the kinetic behavior of the attached myosin motors can be given as
\begin{equation}
	\frac{d n_m}{d t}=\omega_a\left(N_m-n_m\right)-\omega_d^0 n_m \exp \left(\frac{\left|f_l\right|}{f_d}\right).
	\label{nm}
\end{equation}
In their state of attachment, myosin motors exhibit directed movement along the filament bundle, primarily orienting towards the plus end of the filament with a velocity $v_m\left(f_l\right)$, which is dependent on the load force that it experiences.
This behavior is quantitatively captured through a piecewise linear force-velocity relationship \cite{ghosh2022deconstructing,chan2008traction}, formulated as
$$
v_m\left(f_l\right)= \begin{cases}v_u, & \text { for } f_l \leq 0, \\ v_u\left(1-\frac{f_l}{f_s}\right), & \text { for } 0<f_l \leq f_s, \\ v_b, & \text { for } f_l>f_s. \end{cases}
$$
Here, $f_s$ is the stall force, a critical threshold beyond which the myosin motor ceases to move. $v_u$ represents the intrinsic velocity of the motor in the free load, while $v_b$ corresponds to a reverse velocity.

Molecular clutches are modeled as extensible springs with a spring stiffness $\kappa_{c}$, one end attached to the actin bundle and the other to a compliant substrate with stiffness $\kappa_{s}$ through substrate linkage, see Fig.~\ref{myosin_clutch}.
The motion of the myosin motors on the actin bundle induces the retrograde motion of the actin. 
Driven by the contractility of myosin motors, the retrograde motion of the actin bundle with a distance of $x_o$, results in an extension 
$x_c^i$ in the $i$-th clutch among the $n_c$ attached clutches, as well as the substrate extension  $x_s$. 
Let $x_c=\frac{1}{n_c} \sum_{i=1}^{n_c} x_c^i$ represent the average extension of connected clutches. 
There exists a relationship $x_o=x_c+x_s$.
 
Analogous to the mechanism in myosin motors, clutches experience attachment and detachment dynamics at rates $k_{\rm on}$ and $k_{\rm off }$, respectively.
The average tension exerted by attached clutches is $f_l^c=\kappa_c x_c$. 
Correspondingly, the tension along engaged clutches increases the detachment rate, $k_{\rm off}$, exponentially according to Bell model \cite{bell1978models},
$$k_{\rm off}=k_{\rm off}^0 \exp \left(\left|f_l^c\right| / F_b\right),$$ where $k_{\rm off}^{0}$ is the unloaded dissociation rate of the molecular clutch from the actin bundle, $F_b$ is the characteristic bond rupture force.

Defining $N_c$ as the total number of available clutches, we describe the attachment-detachment dynamics of these clutches with the following rate equation
\begin{equation}
	\frac{d n_c}{d t}=k_{\mathrm{on}}\left(N_c-n_c\right)-k_{\mathrm{off}}^0 n_c \exp \left(\frac{\left|f_l^c\right|}{F_b}\right).
	\label{nc}
\end{equation}

The dynamics governing the motion distance $x_o$ of the actin bundle is elucidated by considering a mechanical balance of forces within the over-damped regime, as represented by the following equation:
\begin{equation}
	\Gamma \frac{d x_o}{d t}=n_m \kappa_m y-n_c \kappa_c x_c-F,
	\label{xo}
\end{equation}
where the viscous force  due to the motion of the actin bundle, characterized by the viscous friction coefficient $\Gamma$, is balanced by the collective forces exerted by the motors and clutches, in addition to the external load $F$.

The rate of mean extension of an attached myosin motor is determined by two primary factors: the active velocity of the motor $v_m$ on the actin filament  and the velocity of the actin bundle's movement, represented by  $\frac{d x_o}{dt}$,
\begin{equation}
\frac{d y}{d t}=v_m\left(f_l\right)-\frac{d x_o}{d t}.
\label{y}
\end{equation}
 Since the actin bundle flows in the retrograde direction, there is a negative sign before $\frac{d x_o}{dt}$.

Clutches interact with actin filaments to resist retrograde flow and transmit forces to compliant substrate in the extracellular environment.
In this model, substrate linkage is typically considered dynamic, capable of moving to accommodate changes in internal cellular forces.
The dynamics of the substrate extension $x_s$ are determined by a balance of forces exerted by attached clutches and the substrate, as described by the equation:
\begin{equation}
	\gamma \frac{d x_s}{d t}=n_c \kappa_c x_c-k_s x_s,
	\label{xs}
\end{equation}
where $\gamma \frac{d x_s}{d t}$ represents the substrate frictional damping force due to the motion of the substrate linkage. 
Such a force is akin to that produced by actin filament bundles and is typically quantified by a substrate friction coefficient, $\gamma$.
This relationship signifies that the motion of the substrate linkage engenders resistance proportional to its velocity, mirroring the resistance experienced by objects traversing a viscous fluid. 
Therefore, the substrate friction coefficient, $\gamma$, constitutes a consideration when analyzing the interactions between the motor-clutch model and the substrate.

The extensions of  springs   $x_c$,  $x_s$,  and $y$ are all considered positive when the springs are elongated, and negative during compression, without including the direction sign.

\subsection{Dimensionless Transformation}
These five equations Eqs.~\eqref{nm}--\eqref{xs} constitute a five-dimensional nonlinear autonomous system.
By selecting specific characteristic scales for length, time, velocity, and force as $l_0=\left(k_B T / \omega_d^0 \Gamma\right)^{1 / 2}, \tau=1 / \omega_d^0$, $v_0=l_0 \omega_d^0$, and $f=\left(\omega_d^0 \Gamma k_B T\right)^{1 / 2}$,  respectively, we facilitate the transformation of Eqs.~\eqref{nm}--\eqref{xs} into their dimensionless forms. 
This is accomplished through the introduction of dimensionless parameters: $\tilde{\tau}=t / \tau$,
$\tilde{\omega}=\omega_a / \omega_d^0$, $\tilde{v}_u=v_u / v_0$, $\tilde{k}_{\rm on}=k_{\rm on} / \omega_d^0$, $\tilde{k}_{\rm off }^0=k_{\rm off }^0 / \omega_d^0$, $\tilde{x}_o=x_o / l_0$, $\tilde{x}_s=x_s / l_0$, $\tilde{y}=y / l_0$, $\tilde{\kappa}_c=\kappa_c l_0 / f$, $\tilde{\kappa}_m=\kappa_m l_0 / f$, $\tilde{\kappa}_s=\kappa_s l_0 / f$, $\tilde{F}=F / f$ and $\tilde{f}_s=f_s / f$. 
These dimensionless variables provide a more streamlined and universal framework for analyzing the equations, enhancing the comprehension and applicability of the model under varying physical conditions.
The physical parameters employed in our model are delineated in Tab.~\ref{parameters}.

\begin{table}[ht]
	\centering
	\caption{Physical parameters present in the system}
	\label{parameters}
	\begin{tabular}{l c r}
		\hline
		Parameter &Symbol& Values \\
		\hline
		Motor attachment rate & $\omega_a$ & $40$ ${\rm s}^{-1}$  \cite{walcott2012mechanical}\\
		Motor detachment rate & $\omega_d^0$ & $350$ ${\rm s}^{-1}$  \cite{walcott2012mechanical}\\
		Total number of motors & $N_m$ & 100  \\
		Clutch attachment rate & $k_{\rm on}$ & $1$ ${\rm s}^{-1}$ \cite{chan2008traction}\\
		Clutch detachment rate & $k_{\rm off}^0$ & $0.1$ ${\rm s}^{-1}$  \cite{chan2008traction}\\
		Back velocity & $v_b$ & $0.2256$ $ {\rm \mu m / s}$ \cite{walcott2012mechanical} \\
		Stall force & $f_s$ & $4.96$ pN \cite{walcott2012mechanical}  \\
		Detachment force & $f_d$ & $2.4$ pN  \cite{schnitzer2000force}\\
		Clutch bond rupture force & $F_b$ & $6.25$ pN \cite{bangasser2013determinants}  \\
		Motor spring stiffness & $k_m$ & $0.3$  pN/nm \cite{walcott2012mechanical} \\
		Clutch spring stiffness & $k_c$ & $0.03144$ pN/nm \cite{dembo1988reaction} \\
		Substrate spring stiffness & $k_s$ &  $(10^{-2}\sim 10^2)f/l_{0}$\\
		Viscous friction coefficient & $\Gamma$ & $893\times10^{-6}$ ${\rm  k_B T s/n m}^2$ \cite{lansky2015diffusible}\\
		Substrate friction coefficient & $\gamma$ & $(10^{-5}\sim 10^{-2})\mathrm{k}_{\mathrm{B}} \mathrm{Ts} / \mathrm{nm}^2$ \\
		\hline
		Characteristic length & $l_{0}$ &1.7887 nm  \\
		Characteristic time & $ \tau$ &  0.0029 s\\
		Characteristic velocity & $v_{0}$ & 626.0490 nm/s  \\
		Characteristic force & $f$ &2.3033 $\rm k_B T/$nm \\
		\hline
	\end{tabular}
\end{table}

We now proceed to render our dynamical equations into a dimensionless form:
\begin{equation}
	\begin{aligned}
		\frac{d n_m}{d \tilde{\tau}} & =\tilde{\omega}\left(N_m-n_m\right)-n_m \exp \left(\frac{\tilde{\kappa}_m \tilde{y}}{\tilde{f}_d}\right), \\
		\frac{d n_c}{d \tilde{\tau}} & =\tilde{k}_{\text {on }}\left(N_c-n_c\right)-\tilde{k}_{\text {off }}^0 n_c \exp \left(\frac{\tilde{\kappa}_c \left( \tilde{x}_o-\tilde{x}_s \right) }{\tilde{F}_b}\right),\\
		\frac{d \tilde{x}_o}{d \tilde{\tau}} & =n_m \tilde{\kappa}_m \tilde{y}-n_c \tilde{\kappa}_c\left( \tilde{x}_o-\tilde{x}_s\right)-\tilde{F},\\
		\frac{d \tilde{y}}{d \tilde{\tau}} & =\tilde{v}_u\left(1-\frac{\tilde{\kappa}_m \tilde{y}}{\tilde{f}_s}\right)-\frac{d \tilde{x}_o}{d \tilde{\tau}},\\
		\frac{\gamma}{\Gamma} \frac{d \tilde{x}_s}{d \tilde{\tau}}& =n_c \tilde{\kappa_c} \left( \tilde{x}_o-\tilde{x}_s\right)-k_s \tilde{x}_s.
	\end{aligned}
	\label{Dimensionless_Eq}
\end{equation}
Moreover, we can derive the equation for the average extension of the attached  clutches:
\begin{equation}
	\frac{d \tilde{x}_c}{d \tilde{\tau}}=n_m \tilde{\kappa}_m \tilde{y}-\left( 1+\frac{\Gamma}{\gamma}\right) n_c \tilde{\kappa}_c \tilde{x}_c+k_s \tilde{x}_s-\tilde{F}.
	\label{xc}
\end{equation}
\subsection{Calculation of Fixed Points}

We determine the steady-state solutions, i.e., fixed points of Eqs.~\eqref{Dimensionless_Eq} at steady state, which are represented as follows:

\begin{equation}
	\begin{aligned}
		0& =\tilde{\omega}\left(N_m-n_m\right)-n_m \exp \left(\frac{\tilde{\kappa}_m \tilde{y}}{\tilde{f}_d}\right), \\
		0& =\tilde{k}_{\text {on }}\left(N_c-n_c\right)-\tilde{k}_{\text {off }}^0 n_c \exp \left(\frac{\tilde{\kappa}_c \tilde{x}_c  }{\tilde{F}_b}\right),\\
		0 & =n_m \tilde{\kappa}_m \tilde{y}-n_c \tilde{\kappa}_c \tilde{x}_c-\tilde{F},\\
		0 & =\tilde{v}_u\left(1-\frac{\tilde{\kappa}_m \tilde{y}}{\tilde{f}_s}\right)-0,\\
		0& =n_c \tilde{\kappa}_c \tilde{x}_c-k_s \tilde{x}_s,
	\end{aligned}
	\label{Dimensionless_Eq0}
\end{equation}
where $\tilde{x}_c=\tilde{x}_o-\tilde{x}_s$.

The the fixed points $\tilde{y}^0$, $n_m^0$, $\tilde{x}_s^0$,  $\tilde{x}_o^0$, and $n_c^0$ of the dimensionless dynamical equations are given by:

\begin{align}
		\tilde{y}^0& =\tilde{f}_s / \tilde{\kappa}_m, \label{y0} \\
		n_m^0& =\frac{\tilde{\omega} N_m} {\tilde{\omega}+\exp \left(\tilde{f}_s / \tilde{f}_d\right)}, \label{nm0} \\
		\tilde{x}_s^0& =\frac{ n_m^0 \tilde{f}_s-\tilde{F}}{\tilde{\kappa}_s}, \label{xs0}\\
		\tilde{x}_o^0& =\left( \frac{\tilde{\kappa}_s}{n_c^0\tilde{\kappa}_c}+1\right)\frac{ n_m^0 \tilde{f}_s-\tilde{F}}{\tilde{\kappa}_s}  \label{xo0}.
\end{align}

Subsequently, by the relational expression $\tilde{x}_c=\tilde{x}_o-\tilde{x}_s$, the average  extension of the attached clutches in steady state is

\begin{equation}\tilde{x}_c^0=\frac{ n_m^0 \tilde{f}_s-\tilde{F}}{n_c^0\tilde{\kappa}_c}.
	\label{xc0}
\end{equation}
We substitute $\tilde{x}_c^0$ into the second equation concerning $n_c^0$ in the system of equations Eqs.~\eqref{Dimensionless_Eq0} to obtain
\begin{equation}
	\tilde{k}_{\text {on }}\left(N_c-n_c^0\right)=\tilde{k}_{\text {off }}^0 n_c^0 \exp\left(\frac{n_m^0\tilde{f}_s-\tilde{F}}{n_c^0\tilde{F}_b} \right),
	\label{nc0}
\end{equation}
this is a transcendental equation.
We rearrange Eq.~\eqref{nc0} to obtain
\begin{equation}
	N_c=n_c^0+\frac{\tilde{k}_{\rm off}^0}{\tilde{k}_{\rm on}}n_c^0 \exp\left(\frac{n_m^0\tilde{f}_s-\tilde{F}}{n_c^0\tilde{F}_b} \right), 
	\label{Nc_nc}
\end{equation}
we observe that $N_c$ influences the number of solutions for $n_c^0$. Subsequently, we will investigate the number of solutions for Eq.~\eqref{Nc_nc}. 
Consider the right-hand side of Eq.~\eqref{Nc_nc}, which we denote as $g(n_c^0)$,
\begin{equation}
	g(n_c^0)=n_c^0+\frac{\tilde{k}_{\rm off}^0}{\tilde{k}_{\rm on}}n_c^0 \exp\left(\frac{n_m^0\tilde{f}_s-\tilde{F}}{n_c^0\tilde{F}_b} \right), 
	\label{g(nc)}
\end{equation}
correspondingly, the derivative function $g^\prime(n_c^0)$ is
\begin{equation}
	g^\prime(n_c^0)=1+\frac{\tilde{k}_{\rm off}^0}{\tilde{k}_{\rm on}} \exp\left(\frac{n_m^0\tilde{f}_s-\tilde{F}}{n_c^0\tilde{F}_b} \right)\left(1-\frac{n_m^0\tilde{f}_s-\tilde{F}}{n_c^0\tilde{F}_b} \right).
	\label{derivative}
\end{equation}
It is noted that the sign of $n_m\tilde{f}_s-\tilde{F}$ significantly influences  the positivity or negativity of the derivative function $g^\prime(n_c^0)$. In light of this, we conduct a case-by-case analysis.

{\bf Case 1:} When $n_m^{0}\tilde{f}_s-\tilde{F}>0$, that is $\tilde{F}<n_m^{0}\tilde{f}_s$, the derivative function $g^\prime(n_c^0)$ is monotonically increasing with respect to $n_c^0$. As $n_c^0 \to 0$, $g^\prime(n_c^0)\to -\infty$, and as $n_c^0 \to \infty$, $g^\prime(n_c^0) \to 1+\frac{\tilde{k}_{\rm off}}{\tilde{k}_{\rm on}}$. Therefore, the derivative function $g^\prime(n_c^0)$ undergoes a transition from negative to positive, indicating that the original function $g(n_c^0)$ initially decreases and subsequently increases. 
The minimum value of $N_c$, denoted as $N_{c, {\rm min}}$, corresponds to when $g^\prime(n_c^0)=0$.
Furthermore, Eq.~\eqref{Nc_nc} indicates that the number of solutions for $n_c^0$, can be zero, one, or two, corresponding to the conditions $N_c < N_{c, {\rm min}}$, $N_c=N_{c, {\rm min}}$, and $N_c > N_{c, {\rm min}}$, respectively.

This value of $N_{c, {\rm min}}$ is related to the external load $F$. Within the interval $0 \leq \tilde{F}<n_m^{0}\tilde{f}_s$, $N_{c, {\rm min}}$ monotonically decreases to 0 as $F$ increases.
This point will be further elaborated upon in the subsequent section, as demonstrated in Fig.~\ref{Nc_nc_F}.

Furthermore, we observe that, according to Eq.~\eqref{g(nc)}, $g(n_c^0)$ increases as $n_m^0$ increases.  As $\tilde{\omega}$ increases, it leads to an increase in $n_m^0$. 
When $N_c$ and $\tilde{F}$ are held constants, there exists a maximum allowable value for $\tilde{\omega}$, denoted as $\tilde{\omega}_{\rm max}(N_c, \tilde{F})$, such that when $\tilde{\omega}>\tilde{\omega}_{\rm max}(N_c, \tilde{F})$, the minimum value of $g(n_c^0)$  exceeds the $N_{c}$,  i.e., $g(n_c^0)_\text{min}> N_{c}$. 
As a result, no solutions exist for Eq.~\eqref{Nc_nc}. 
It is easy to observe that $\tilde{\omega}_{\rm max}(N_c, \tilde{F})$ is dependent on $N_c$ and increases as $N_c$ increases.

{\bf Case 2:} When $n_m^{0}\tilde{f}_s-\tilde{F}\le 0$, that is $\tilde{F}\ge n_m^{0}\tilde{f}_s$, the derivative function $g^\prime(n_c^0)$ is strictly positive. This results in the original function $g(n_c^0)$ being monotonically increasing with respect to $n_c^0$. Notably, when $n_c^0\to 0$,  $g(n_c^0)\to 0^{+}$. Therefore, Eq.~\eqref{Nc_nc} has only the trivial solution where $N_c=n_c^0=0$. 
This means that the external force $F$ balances the force generated by the myosin motors, rendering the clutches unnecessary.

\subsection{Linear Approximation of Nonlinear Systems}
We rewrite five-dimensional nonlinear system Eqs.~\eqref{Dimensionless_Eq} in vector form,

\begin{equation}
	\frac{d}{d \tilde{\tau}}\left(\begin{array}{c}
		n_m \\
		n_c\\
		\tilde{x}_o \\
		\tilde{y} \\
		\tilde{x}_s
	\end{array}\right)=\left(\begin{array}{c}
		\tilde{\omega}\left(N_m-n_m\right)-n_m \exp \left(\frac{\tilde{\kappa}_m \tilde{y}}{\tilde{f}_d}\right)\\
		\tilde{k}_{\text {on }}\left(N_c-n_c\right)-\tilde{k}_{\text {off }} n_c \exp \left(\frac{\tilde{\kappa}_c  \tilde{x}_c  }{\tilde{F}_b}\right)\\
		n_m \tilde{\kappa}_m \tilde{y}-n_c \tilde{\kappa}_c  \tilde{x}_c -\tilde{F} \\
		\tilde{v}_u\left(1-\frac{\tilde{\kappa}_m \tilde{y}}{\tilde{f}_s}\right)-\left(n_m \tilde{\kappa}_m \tilde{y}-n_c \tilde{\kappa}_c \tilde{x}_c-\tilde{F} \right) \\
		\frac{\Gamma}{\gamma}\left(  n_c \tilde{\kappa_c} \tilde{x}_c-k_s \tilde{x}_s\right) 
	\end{array}\right),
	\label{vector}
\end{equation}
where $\tilde{x}_c=\tilde{x}_o-\tilde{x}_s$.

To analyze the stability near fixed points, we perform a linear approximation of this five-dimensional nonlinear autonomous system.
Below is the linear system we introduce. To distinguish it from the preceding nonlinear system, we rename the variables as $\delta n_m, \delta n_c, \delta \tilde{x}_o, \delta \tilde{y}, \delta \tilde{x}_s$,
\begin{equation}
	\frac{d}{d \tilde{\tau}}\left(\begin{array}{c}
		\delta n_m \\
		\delta n_c\\
		\delta \tilde{x}_o \\
		\delta \tilde{y} \\
		\delta \tilde{x}_s
	\end{array}\right)=\mathcal{J}\left(\begin{array}{c}
		\delta n_m \\
		\delta n_c\\
		\delta \tilde{x}_o \\
		\delta \tilde{y} \\
		\delta \tilde{x}_s
	\end{array}\right),
\end{equation}
where $\mathcal{J}$ is  the $5\times5$ Jacobian matrix 
 of the fixed points, as detailed in Appendix.
The eigenvalues of this Jacobian matrix $\mathcal{J}$ determine the linear stability of the dynamical system.
The determinant of the Jacobian matrix 
$\mathcal{J}$  has been computed as follows:

\begin{align}
	|\mathcal{J}| = &-\frac{\tilde{v}_u\tilde{\kappa}_{c} \tilde{\kappa}_{m} \tilde{\kappa}_{s}}{\tilde{F}_{b} \tilde{f}_{s}}
	\exp\left( \frac{-\tilde{F}}{n_{c}^0 \tilde{F}_{b}}\right) \left( \exp\left( \frac{\tilde{f}_s}{\tilde{f}_d}\right)  + \tilde{\omega}\right) \nonumber\\
	&\times  \left[
	\tilde{k}_{\rm off} \exp\left( \frac{n_{m}^{0}\tilde{f}_{s}}{n_{c}^0\tilde{F}_b}
	\right)  \left( \tilde{F} + n_{c}^0 \tilde{F}_b - n_{m}^0 \tilde{f}_s\right)
	\right. + \\
	& \left. 	\quad \;\, \exp\left( \frac{\tilde{F}}{n_{c}^0 \tilde{F}_b}\right)  n_{c}^0 \tilde{F}_b \tilde{k}_{\rm on} 
	\right]. \nonumber
\end{align}
The characteristic polynomial of the Jacobian matrix is expressed as
\begin{equation}
	P(\lambda)=(\lambda-\lambda_1)( \lambda-\lambda_2)( \lambda-\lambda_3)( \lambda-\lambda_4)( \lambda-\lambda_5) ,
\end{equation}
which forms a fifth-order polynomial equation, with $\lambda_1, \cdots, \lambda_5$ as its eigenvalues.

\section{Results}

\subsection{Saddle-Node Bifurcation: Stable and Unstable Branches}

As previously mentioned, when $N_{c}<N_{c, {\rm min}}$, fixed points do not exist in the dynamical system.
However, as $N_{c}$ increases and reaches $N_{c, {\rm min}}$, the system undergoes bifurcation: a saddle-node bifurcation transforms a single saddle-node into two fixed points with differing stabilities.
One of these is a stable fixed point, forming a stable branch, while the other is an unstable fixed point, giving rise to an unstable branch. 
By analyzing the determinant of the Jacobian matrix, we can readily identify certain unstable fixed points.
This is due to the fact that the determinant of the Jacobian matrix is equivalent to the product of all its eigenvalues, represented as $|\mathcal{J}| =\lambda_1 \lambda_2\lambda_3\lambda_4\lambda_5$. 
If $|\mathcal{J}|>0$, it implies the existence of at least one positive eigenvalue, indicating that the system is unstable.

Upon computation, it has been confirmed that for the smaller fixed point, substituting $n_{c}^{0}$ into the Jacobian matrix yields $|\mathcal{J}|>0$, implying the existence of at least one positive eigenvalue and indicating that the system is unstable. 
Consequently, our research focus is redirected towards the stable branch, where $|\mathcal{J}|<0$.

\begin{figure}[htbp]
	\includegraphics[scale=0.35]{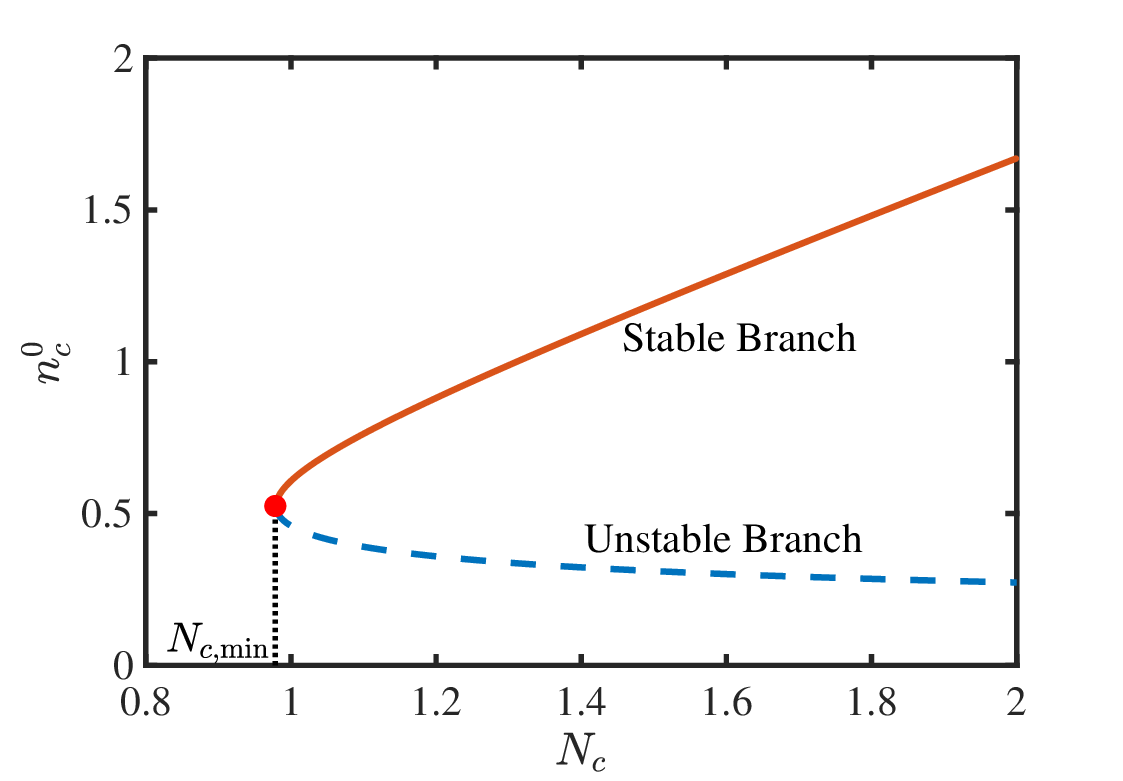}\\
	\caption{\label{nc_Nc}
		A saddle-node bifurcation emerges from the transcendental properties of the $n_{c}^{0}$ equation.
		The solid curve represents the stable branch, while the dashed line indicates the unstable branch. The highlighted point marks the bifurcation point where the system's behavior changes.
		It is noteworthy that $N_{c, {\rm min}}=0.978$ represents the critical threshold for the existence of solutions. In this figure, the external load $\tilde{F}=0$.}
\end{figure}

The bifurcation point can be obtained through the condition where the derivative in Eq.~\eqref{derivative} equals zero, $g^\prime(n_c^0)=0$, in conjunction with Eq.~\eqref{nc0}.
We ascertain the bifurcation point and mark it in Fig.~\ref{nc_Nc} as a highlighted point.

In the region where $N_{c}<N_{c, {\rm min}}$ is lower than its value at the bifurcation point, due to the absence of fixed points for $n^{0}_{c}$, the system loses stability and remains unstable regardless of $\tilde{v}_{u}$. Consequently, this region is labeled as the Non-existence (N) zone in the stability diagram shown in Fig.~\ref{Nc_vu}.

\subsection{Different Dynamical Behavior of the System}

When $|\mathcal{J}|<0$, the dynamical behavior of the system, which has four different states characterized by the different combinations of the five eigenvalues: 

(1) Five real negative eigenvalues are indicative of a stable (S) state, wherein perturbations exhibit exponential decay over time.

(2) There are three real negative and two real positive eigenvalues; alternatively, there is one real negative and four real positive eigenvalues.
Both configurations give rise to an unstable (U) state characterized by exponentially intensifying perturbations.

(3) There are three real negative eigenvalues and two complex conjugate eigenvalues with a negative real part, denoted as $\lambda_{4,5}=-\alpha \pm i \beta$. This configuration characterizes a stable spiral (SS) state with decaying oscillations.

(4) There are three real negative eigenvalues and two complex conjugate eigenvalues with a positive real part, expressed as $\lambda_{4,5}=\alpha \pm i \beta$.
This represents an unstable spiral (US) state with oscillations of growing amplitude.

\begin{table}[b]
	\caption{\label{tab:1} Different dynamical states are associated with combinations of eigenvalues. In this table, $\alpha$ is a positive real number. The numbers in this table correspond to the quantities of the respective types of eigenvalues.}
	\begin{ruledtabular}
		\begin{tabular}{ccccc}
			& $\alpha$ & $-\alpha$ & $-\alpha \pm i\beta$ & $\alpha \pm i\beta$ \\
			\colrule
			S (Stable) & 0 & 5 & 0 & 0 \\
			\hline
			SS (Stable Spirals) & 0 & 3 & 2 & 0 \\
			\hline
			US (Unstable Spirals) & 0 & 3 & 0 & 2 \\
			\hline
			\multirow{2}{*}{U (Unstable)} & 2 & 3 & 0 & 0 \\
			& 1 & 4 & 0 & 0 \\
		\end{tabular}
	\end{ruledtabular}
	\raggedright
\end{table}

In the system under consideration, as the parameter values change, there emerges an increasing oscillation over time transitioning from a decaying oscillation.
This phenomenon represents a dynamical transition from a stable spiral (SS) to an unstable spiral (US), attributed to a change in the complex conjugate eigenvalues from $(-\alpha \pm i \beta)$ to $(\alpha \pm i \beta)$.
The sign of the real part of the complex conjugate roots $\alpha$ is opposite on either side of the boundary; consequently, $\alpha=0$ is the condition of the associated phase boundary.
This transition from SS to US is the route through which stable limit cycle oscillation sets in the system via non-linear effects leading to Hopf bifurcation. For theoretical methods related to the stability of limit cycles and the assessment of Hopf bifurcations, see reference \cite{verduzco2005first}.

In Fig.~\ref{validate}, we provide visual representations of the four dynamical behaviors exhibited by the system. In the S state, the system rapidly stabilizes, reaching steady-state solutions.
Conversely, in the SS state, the system undergoes a trajectory of decaying oscillations.
Notably, within the US state, the system exhibits self-sustaining limit cycle oscillations, a phenomenon observed subsequent to crossing the supercritical Hopf bifurcation boundary.
In the U state, the variables $\tilde{x}_o$, $\tilde{y}$, and $\tilde{x}_s$ diverge, failing to converge to a fixed point. 

Furthermore, we present trajectory diagrams within the $n_m-\tilde{x}_0$ and $\tilde{y}-\tilde{x}_0$ planes under  SS and US states, as shown in Fig.~\ref{Trajectory}.
In the SS state, the trajectories eventually converge towards a fixed point, whereas in the US state, the  trajectories ultimately evolve towards and settle on a limit cycle, indicating a persistent oscillatory behavior.

\subsection{Linear Stability Analysis of the System}
\begin{figure}[htbp]
	\includegraphics[scale=0.37]{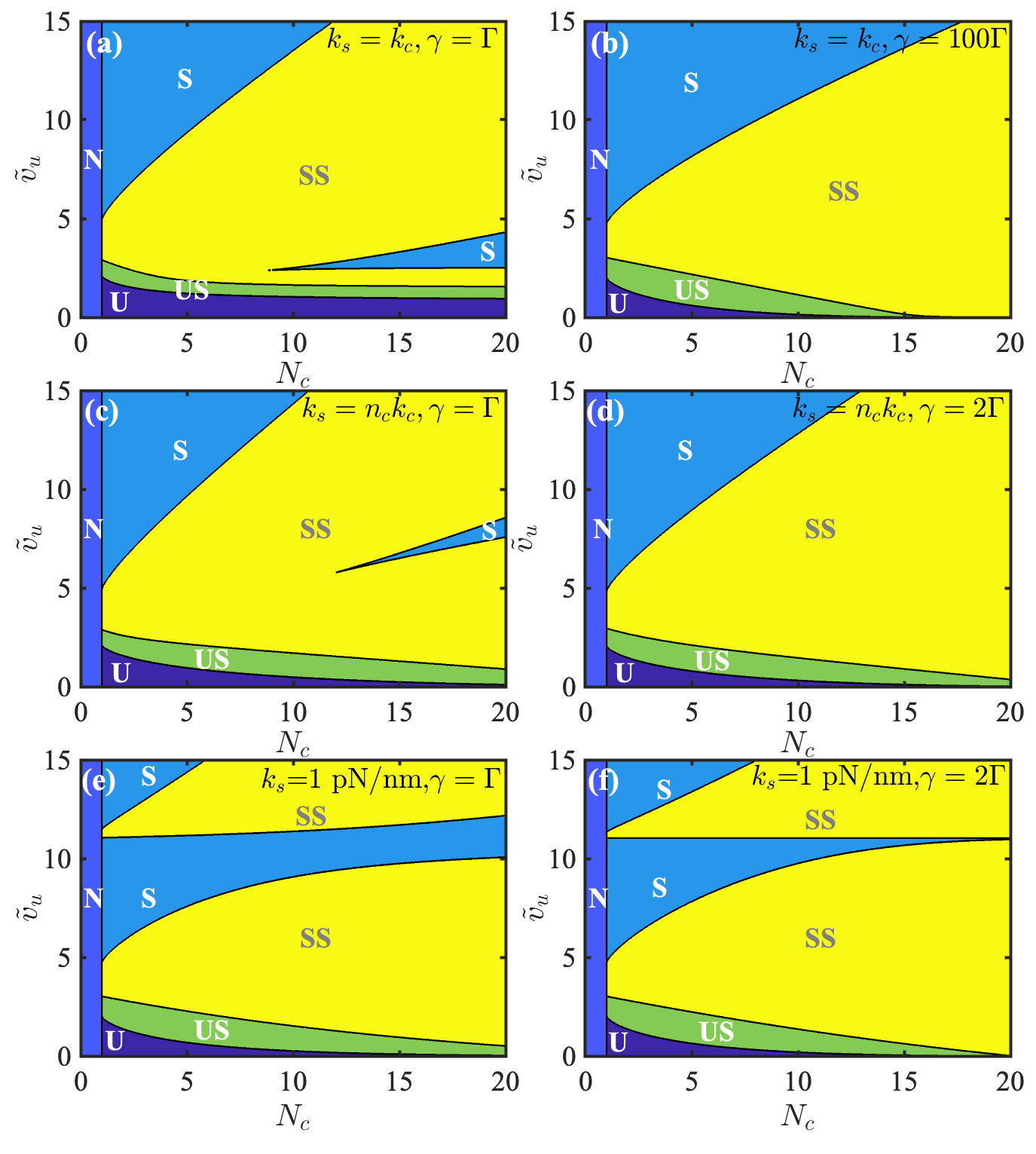}\\
	\caption{\label{Nc_vu} Stability diagram of the  $N_{c}-\tilde{v}_{u}$ plane for the system.
	Regions marked with different letters represent distinct dynamical behaviors: \lq S' represents  stable , \lq SS' indicates stable spirals, \lq US' designates unstable spirals, and \lq U' corresponds to unstable. The region marked \lq N' signifies the non-existence of fixed points.
Each dynamical stability region corresponds to a combination of eigenvalues. The stability at each point is determined by calculating the eigenvalues of the Jacobian matrix for a linearly approximated system, varying with changes in $N_{c}$ and  $\tilde{v}_{u}$. In this study, we utilized an extensive range for $\tilde{v}_{u}$, spanning from 0 to 15.  The scope of our research on $N_{c}$ encompasses the range $0-20$. The phase boundaries lack specific analytical forms and are derived through numerical computations.  
The theoretical predictions are rigorously assessed through numerical solutions of the differential equations across the full spectrum of the parameter plane.}
\end{figure}

In this section, we explore the stability of the system when it is unloaded, namely when the external load $\tilde{F}=0$. We analyze the impact of relevant parameters on the stability of this dynamic system from several perspectives: myosin motors, molecular clutches, and the compliant substrate. The parameters characterizing myosin Il activity include the velocity of attached myosin motors and their attachment and detachment rates with the actin filament.

Given that $N_c$ controls the saddle-node bifurcation and $\tilde{v}_u$ regulates the activity of the myosin motors, we present a stability diagram in the $N_c-\tilde{v}_u$ plane, which delineates the various dynamical behaviors exhibited by our system.
Additionally, we use the model to reproduce the results of the $N_{c}-\tilde{v}_{u}$ plane stability diagram for the dynamic system observed under various experimental conditions, namely different substrate stiffnesses and different substrate viscous friction coefficients, as shown in Fig.~\ref{Nc_vu}.

From Fig.~\ref{nc_Nc}, within the stable branch, a smaller $N_{c}$ corresponds to a smaller $n_{c}^{0}$, which leads to a larger equilibrium position $\tilde{x}_{c}^{0}$ (as deduced from Eq.~\eqref{xc0}). 
Consequently, the detachment rate $\tilde{k}_{\mathrm{off}}$ increases, as this is due to $\tilde{k}_{\rm off}=\tilde{k}_{\rm off}^0 \exp \left(\frac{\tilde{\kappa}_c  \tilde{x}_c  }{\tilde{F}_b}\right)$.
When $N_{c}<N_{c,{\rm min}}$ and $N_{c}$ continues to decrease, $\tilde{k}_{\mathrm{off}}$ further increases, causing $\frac{d n_c}{d \tilde{\tau}}<0$ to always hold true, corresponding to no solution for Eq.~\eqref{nc0}. 
This results in $n_c \to 0$, and to balance the forces of myosin motors and the clutches, $\tilde{x}_c \to \infty$, which is equivalent to the clutch mechanism being inoperative and unable to resist the retrograde motion of the actin.
This region is marked as \lq N' in Fig.~\ref{Nc_vu}.

Earlier experimental investigations \cite{harris1993smooth,debold2005slip,debold2011phosphate} have established the force-velocity relationships of myosin and performed ensemble measurements of motor velocities under unloaded conditions.
As ATP concentration increases,  motor velocities $\tilde{v}_{u}$ increase correspondingly, the system more likely to enter a stable state. This phenomenon elucidates the system's propensity to attain a stable state at high ATP concentrations.

At high motor velocities, the system maintains a stable state. However, with a decrease in motor velocity, the system transitions into a stable spiral state.
In the SS region, the oscillations gradually decay with the increase in time.
As $\tilde{v}_{u}$ progressively decreases, the system crosses into US region of periodic oscillations.  
In the US region, the system exhibits a limit cycle around an unstable fixed point.
As $\tilde{v}_{u}$ decreases, the transition from a stable fixed point to oscillatory behavior (represented by the emergence of a stable limit cycle from an unstable fixed point) is a hallmark of a supercritical Hopf bifurcation. 
Additionally, when motor velocity $\tilde{v}_{u}$ is very low, the system transitions from U region to US region upon an increase in $N_{c}$.

Considering that cells utilize the motor-clutch mechanism to probe their mechanical environment, we wonder how this system would be influenced by substrate stiffness. When we simulate the motor-clutch system on compliant substrates, we classify the substrate stiffness into three distinct categories based on its relationship with the ensemble clutch stiffness.
Firstly, when the substrate stiffness $k_{s}=k_{c}$ and is found to be less than the ensemble clutch stiffness, the condition is defined as a \lq soft substrate'.  
Secondly, a \lq stiff substrate' condition occurs when $k_{s}=1$ pN/nm, with the substrate stiffness exceeding that of the ensemble clutch stiffness. The maximum value of the ensemble clutch stiffness, $n_{c}k_{c}$, remains below 1 pN/nm, even as $n_{c}$ increases in response to variations in $N_{c}$. Finally, the substrate stiffness $k_{s}$ is equivalent to the number of engaged clutches multiplied by the individual clutch stiffness, namely $k_{s}=n_{c}k_{c}$, which we define as a \lq matched substrate'. In this context, the ensemble clutch stiffness precisely matches the substrate stiffness, indicating that the ensemble clutch and substrate, both having equal stiffness, are in series.  Consequently, the overall stiffness becomes $\frac{1}{2}n_{c}k_{c}$. 

If the substrate is too soft, cells may not be able to effectively transmit forces, leading to compromised cell motility and signal transduction. On the other hand, if the substrate is too hard, while force transmission can be effective, it might negatively affect the cell's flexibility and adaptability. 
A majority of theoretical studies \cite{elosegui2016mechanical} focusing on force transmission within clutch models, wherein substrate stiffness is adjusted, reveal a biphasic relationship between the stiffness of the substrate and the clutches.
This biphasic correlation between the stiffness of the substrate and that of the clutches implies the existence of an optimal range of stiffness, within which cellular forces are transmitted most efficiently. 
Both excessively rigid or overly soft substrates may reduce the efficiency of force transmission.

From Fig.~\ref{Nc_vu} (a), (c), and (e),  we explore the influence of varying substrate stiffness on system stability.
Our findings indicate that, for a soft substrate as shown in Fig.~\ref{Nc_vu} (a), it is notably difficult to achieve a transition from U to US by increasing $N_{c}$ at lower velocities $0<\tilde{v}_{u}<1$.
Conversely, with an increase in substrate stiffness, as shown in Fig.~\ref{Nc_vu} (c) and (e), we observe that at similar lower velocities $\tilde{v}_{u}$, the system can transition from U to US, provided a sufficient number of clutches are available, by increasing $N_{c}$.
With the increase in substrate stiffness $\tilde{k}_s$, the upper boundary of US becomes progressively steeper and more inclined. Concurrently, the area of US region expands, while U region diminishes. This elucidates that an increase in substrate stiffness leads to an expansion in the total area of the S and SS regions, thereby exerting a stabilizing effect on the system.

In terms of effects, increasing the substrate friction coefficient and the enhancement of substrate stiffness both act to inhibit the movement of the substrate linkage.
Comparing Fig.~\ref{Nc_vu} (a) and (b), (c) and (d), as well as (e) and (f), it can be observed that increasing substrate friction coefficient also exerts a stabilizing influence on the system.
Particularly, Fig.~\ref{Nc_vu}(b) employs a considerably large substrate viscous friction coefficient to simulate a condition wherein the substrate linkage is virtually immobilized. This is akin to the assumptions of the model in \cite{ghosh2022deconstructing}, resulting in a similar stability distribution diagram.

\begin{figure}[htbp]
	\includegraphics[scale=0.35]{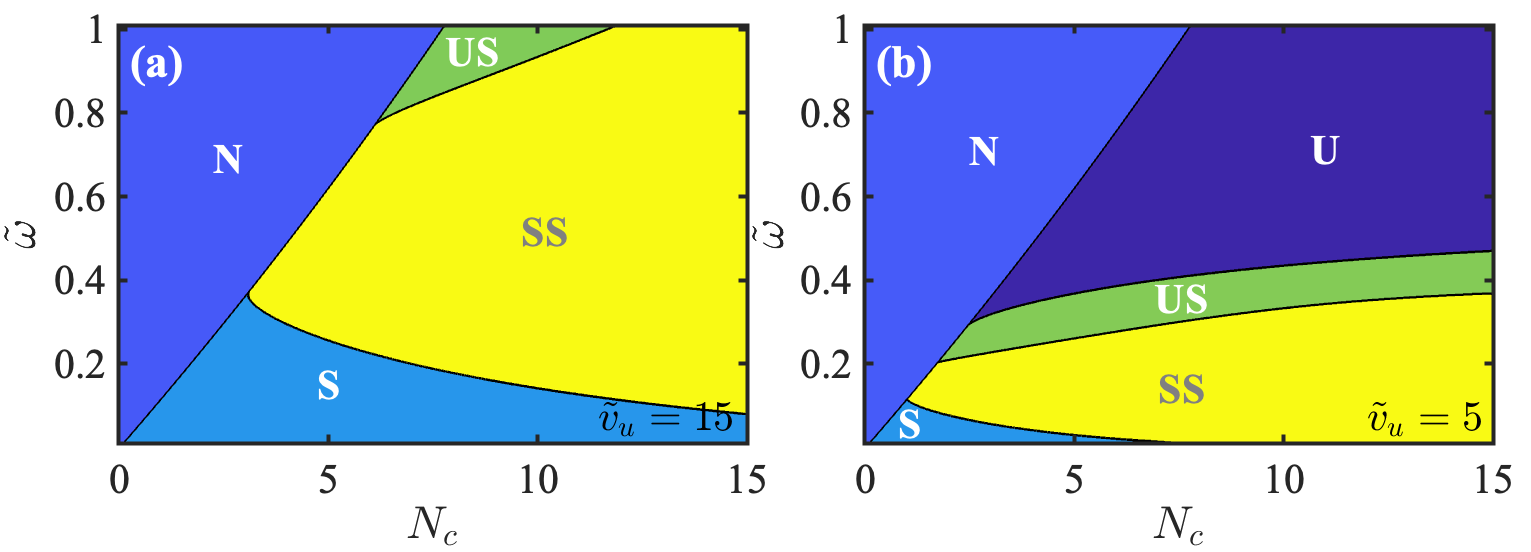}\\
	\caption{\label{omega_Nc}Stability diagram of the $N_{c}-\tilde{\omega}$ plane. The distinction between (a) and (b) is the activity of the myosin motors; specifically, the motors exhibit a higher velocity in (a), with $\tilde{v}_{u}=15$, while in (b), the motor speed is lower, $\tilde{v}_{u}=5$. }
\end{figure}
Next, we elucidate the impact of the ratio $\tilde{\omega}$ on system stability by investigating the $N_{c}-\tilde{\omega}$ plane stability diagram.
In Fig.~\ref{omega_Nc}, with the constant $N_c$, according to Eq.~\eqref{nm0}, we observe that an increase in $\tilde{\omega}$ leads to an increase in $n_{m}^{0}$, the force exerted by myosin  $n_{m}\tilde{k}_m \tilde{y}$ increases, causing $\frac{d \tilde{x}_c}{d \tilde{\tau}}>0$ in Eq.~\eqref{xc} and resulting in an increase in $\tilde{x}_c$. This change is a response by the clutch mechanism to achieve equilibrium, implying that the force applied by the clutch should also increase. 

Moreover, the detachment rate $\tilde{k}_{\rm off}$ increases accordingly. 
As $\tilde{\omega}$ exceeds $\tilde{\omega}_{\rm max}(N_c, \tilde{F})$, $\frac{d n_c}{d \tilde{\tau}}<0$ consistently holds, corresponding to the absence of solutions for Eq.~\eqref{nc0}. This leads to $n_c \to 0$ and $\tilde{x}_c\to \infty$. The fixed point vanishes, and the system enters the N region. Except for the N region, when $N_c$ is fixed, an increase in $\tilde{\omega}$ leads to the system becoming more unstable. This indicates that an increase in $\tilde{\omega}$ results in a destabilizing effect on the system. 
 
 In Fig.~\ref{omega_Nc}(a)(b), as $\tilde{\omega}$ increases, a Hopf bifurcation is observed during the transition from the SS state to the US state.

\begin{figure}[htbp]
	\includegraphics[scale=0.35]{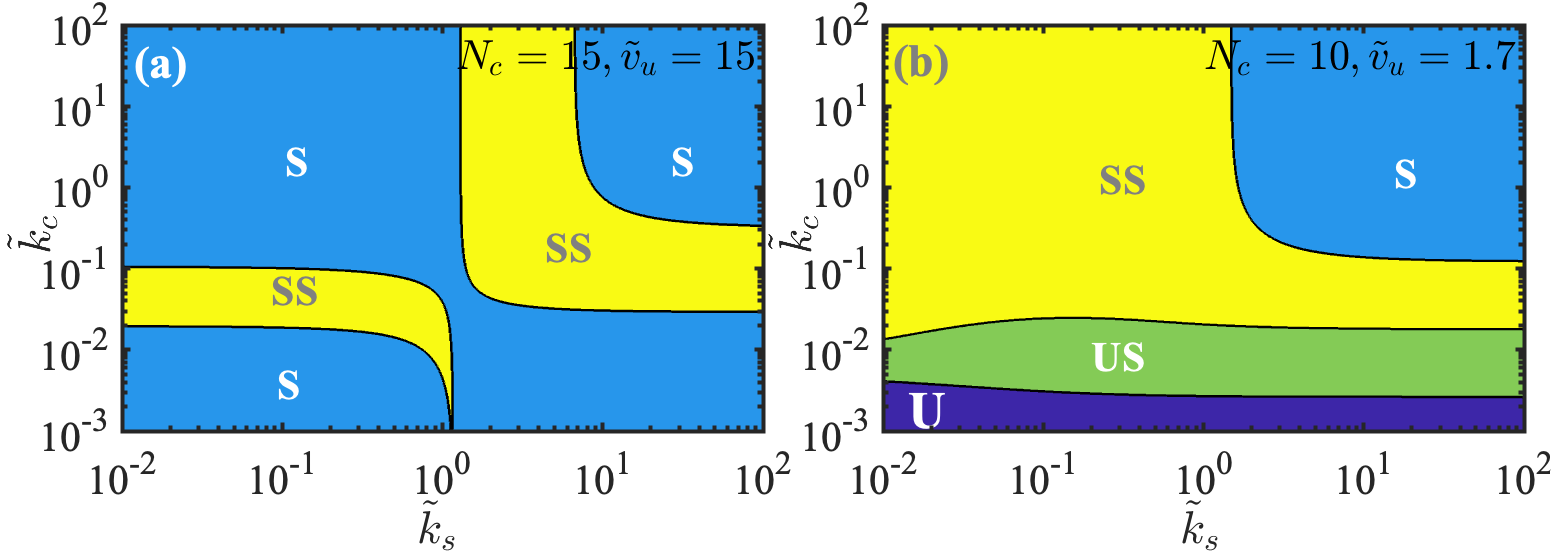}\\
	\caption{\label{kc_ks}Stability diagram of the $\tilde{k}_{s}-\tilde{k}_{c}$ plane. In (a), both $N_c$ and $\tilde{v}_u$ are significantly large. Conversely, in (b), $N_c$ is set at 10 and $\tilde{v}_u$ at 1.7, positioning this point close to the upper boundary of the US region in Fig.~\ref{Nc_vu}(a), where $\tilde{v}_u$ is comparatively lower.}
\end{figure}

As shown in Fig.~\ref{kc_ks}, we investigate the impact of the clutch spring stiffness $k_c$ on the system's stability through the stability diagram in the $\tilde{k}_{s}-\tilde{k}_{c}$ plane.
Within the research scope depicted in Fig.~\ref{kc_ks}(a), when both $N_c$ and $\tilde{v}_u$ are sufficiently large, the system remains stable regardless of variations in $\tilde{k}_s$ and $\tilde{k}_c$; it simply transitions back and forth between the S and SS regions. 
In Fig.~\ref{kc_ks}(b), $N_c=10$ and $\tilde{v}_u=1.7$, corresponding to a point near the SS and US boundary in Fig.~\ref{Nc_vu}(a), where $\tilde{v}_u$ is relatively low.
We find that the clutch spring stiffness, $\tilde{k}_c$, should not be too low. When the clutch stiffness is extremely low, the system remains unstable or enters a state of unstable spirals, regardless of variations in  $\tilde{k}_s$. 
In Fig.~\ref{kc_ks}(b), when $\tilde{k}_s$ is fixed, an increase in $\tilde{k}_c$ enhances the propensity of the system towards stabilization.
Consequently, $\tilde{k}_c$ plays a pivotal role in the stability of the system.

\begin{figure}[htbp]
	\includegraphics[scale=0.34]{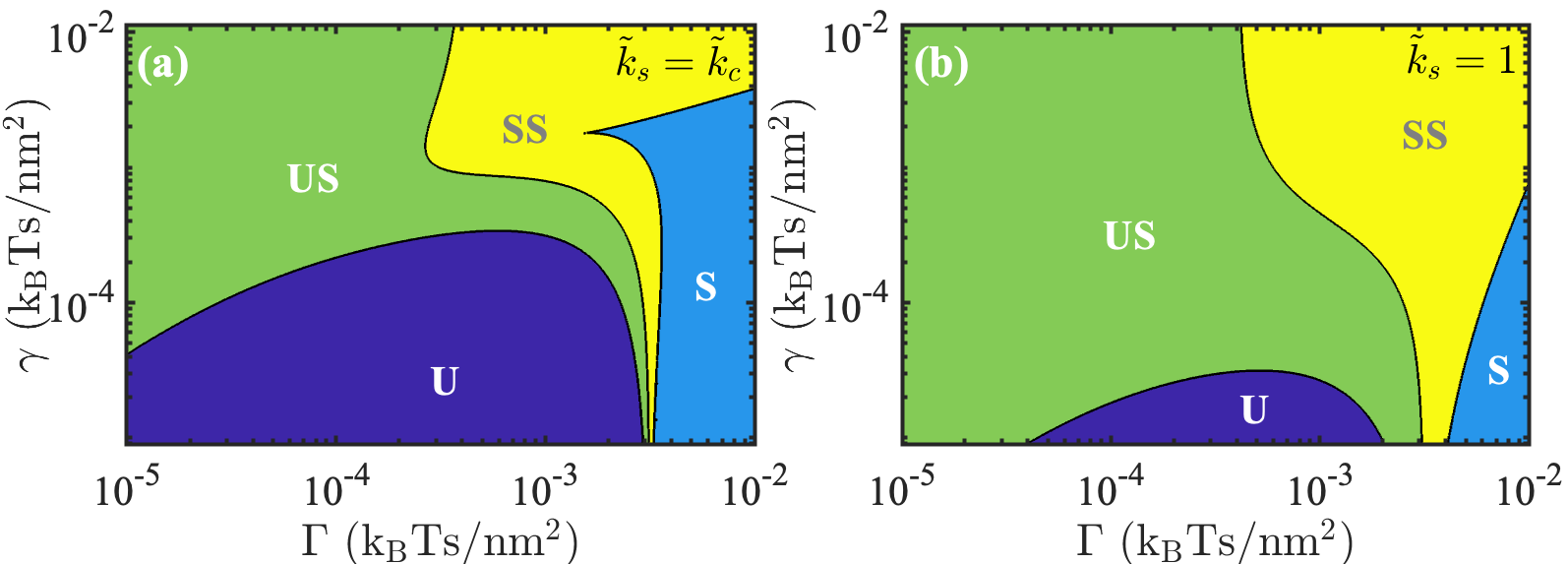}\\
	\caption{\label{alpha_gamma}Stability diagram of the $\Gamma-\gamma$ plane. The difference between the two pictures (a) and (b), is the stiffness of the substrate; Fig.~\ref{alpha_gamma}(a) features a soft substrate, while Fig.~\ref{alpha_gamma}(b) features a stiff substrate.  }
\end{figure}
In Fig.~\ref{alpha_gamma}, we investigate the impact of two viscous friction coefficients, $\Gamma$ and $\gamma$, on the stability of the system.
Upon comparing Fig.\ref{alpha_gamma}(a) and Fig.\ref{alpha_gamma}(b), it is evident that an increase in substrate stiffness $\tilde{k}_s$ leads to a reduction in the U region.
As the viscous friction coefficient $\Gamma$ and the substrate friction coefficient $\gamma$ decrease, the system becomes unstable or oscillatory, corresponding to the lower left region in the Fig.~\ref{alpha_gamma}.
Particularly, when $\Gamma$ is significantly low, that is, less than $10^{-4}$ $\mathrm{k}_{\mathrm{B}} \mathrm{Ts} / \mathrm{nm}^2$, the system becomes unstable or exhibits unstable spirals, regardless of the variations in $\gamma$.

\subsection{The Impact of Parameters on Fixed Points} 
Due to the complexity of regulating force generation and the subsequent transmission process, understanding the role of each parameter in this process, as well as how they interact with one another, is crucial for a deep understanding of cellular mechanical behavior. 
Additionally, mechanical force measurements vary significantly at the cellular level \cite{kurzawa2017dissipation,meili2010myosin,rape2011regulation,plotnikov2012force}.
Therefore, there is a need to discern the role of individual parameters on fixed points.
Based on the expressions of the fixed points Eq.~\eqref{y0}-\eqref{xc0}, we can analyze the influence of certain parameters on the fixed points.
However, since the influence of some parameters cannot be intuitively analyzed through the analytical expressions, we also plotted the corresponding numerical trend graphs, as shown in Figs.~\ref{nc}, ~\ref{xo}, and ~\ref{fs}. 
Combining the above two methods, we ultimately summarize the influence of each parameter on the fixed points in  Tab.~\ref{influence}.

\begin{table}[b]
\caption{	\label{influence}
The influence of parameters on fixed points is depicted in the table.  An upward arrow ($\uparrow$) indicates a monotonic increase, a downward arrow ($\downarrow$) signifies a monotonic decrease, and a combination of an upward followed by a downward arrow ($\uparrow\downarrow$) represents an influence that initially increases and then decreases.  Conversely, a downward followed by an upward arrow ($\downarrow\uparrow$) would represent an influence that initially decreases and then increases.  A dash (--) indicates no influence. In this table, monotonicities derived directly from analytical expressions are represented by single arrows, whereas those determined through numerical plotting are denoted by double arrows to distinguish between the two methodologies.}
	\begin{ruledtabular}
		\begin{tabular}{@{}ccccccc@{}}
		 & $n_{m}^{0}$ & $n_{c}^{0}$ & $\tilde{x}_{o}^{0}$ & $\tilde{y}^{0}$ &  $\tilde{x}_{s}^{0}$ &  $\tilde{x}_{c}^{0}$ \\ \hline
		$\tilde{\omega}$ & $\uparrow$ & $\Downarrow$ & $\Uparrow$ & -- &$\uparrow$ & $\Uparrow$ \\
		$N_m$& $\downarrow$ & $\Downarrow$ &  $\Uparrow$& -- & $\uparrow$ & $\Uparrow$ \\
		$N_c$& --  & $\Uparrow$ & $\Downarrow$ & -- & -- & $\Downarrow$ \\
		$\tilde{k}_{\rm off}/\tilde{k}_{\rm on}$& -- & $\Downarrow$ &  $\Uparrow$ & -- & -- &  $\Uparrow$ \\
		$\tilde{f}_s$& $\downarrow$ & $\Downarrow\Uparrow$ & $\Uparrow\Downarrow$ & $\uparrow$ &$\uparrow\downarrow$& $\Uparrow\Downarrow$ \\
		$\tilde{f}_d$& $\uparrow$ & $\Downarrow$ & $\Uparrow$ & -- &$\uparrow$ & $\Uparrow$ \\
		$\tilde{F}_b$& -- & $\Uparrow$ & $\Downarrow$ & -- & -- & $\Downarrow$ \\
		$\tilde{\kappa}_m$& -- & --&  --& $\downarrow$& -- & -- \\
		$\tilde{\kappa}_c$& -- & -- & $\downarrow$ & -- & -- & $\downarrow$ \\
		$\tilde{\kappa}_s$& -- & -- & $\downarrow$ & -- & $\downarrow$ & -- \\
		$\tilde{F}$&  -- & $\uparrow$ & $\downarrow$ & -- & $\downarrow$ &$\downarrow$  \\
	\end{tabular}
\end{ruledtabular}
\end{table}

Specifically, the direct impact of various parameters on $n_c^{0}$, as suggested by Eq.\eqref{nc0}, presents a challenge for intuitive analysis through analytical means alone. To address this, we offer detailed trend graphs in Fig.~\ref{nc}, providing a visual representation of these parameter influences. 
As depicted in Fig.~\ref{nc}(a), we observe that $n_c^{0}$ increases with $N_c$.
When $N_c<11.92$, the system is in state S;  when $N_c$ exceeds 11.92, it enters the SS state. With a sufficient number of clutches available, the system transitions from state $\mathrm{S}$ to $\mathrm{SS}$.

In Fig.~\ref{nc}(b), $n_c^{0}$ decreases slowly as $N_m$ increases. 
For $N_c=10$, the system is in the $\mathrm{S}$ state when $N_m$ is below 123; it shifts to the SS state as $N_m$ falls between 123 and 150. 
When $N_c=20, N_m$ is consistently in state SS within the range $[50,150]$.

Furthermore, in Fig.~\ref{nc}(c), $n_c^{0}$ monotonically decreases with an increase in $\tilde{\omega}$.
For $N_c=10$, the system is in state S when $\tilde{\omega}<0.14$, transitioning to state SS beyond that threshold.
Similarly, for $N_c=20$, the S state is observed when $\tilde{\omega}$ is under 0.05 , with the SS state emerging thereafter.
In Fig.~\ref{nc}(d), $n_c^{0}$ monotonically decreases with an increase in $k_{\rm off}/k_{\rm on}$. For $N_c=10$, the system remains in the S state across the entire range of $[0,1]$ for $\tilde{k}_{\rm off}/\tilde{k}_{\rm on}$. At $N_c=20$, it is in the SS state when $\tilde{k}_{\rm off}/\tilde{k}_{\rm on}$ is below 0.78, and transitions to the S state for values between 0.78 and 1. 
Continuing to Fig.~\ref{nc}(e), $n_c^{0}$ increases with an increase in $\tilde{F}_b$.  
Lastly, in Fig.~\ref{nc}(f), $n_c^{0}$ gradually decreases as $\tilde{f}_d$ increases.

\begin{figure}[htbp]
	\includegraphics[scale=0.34]{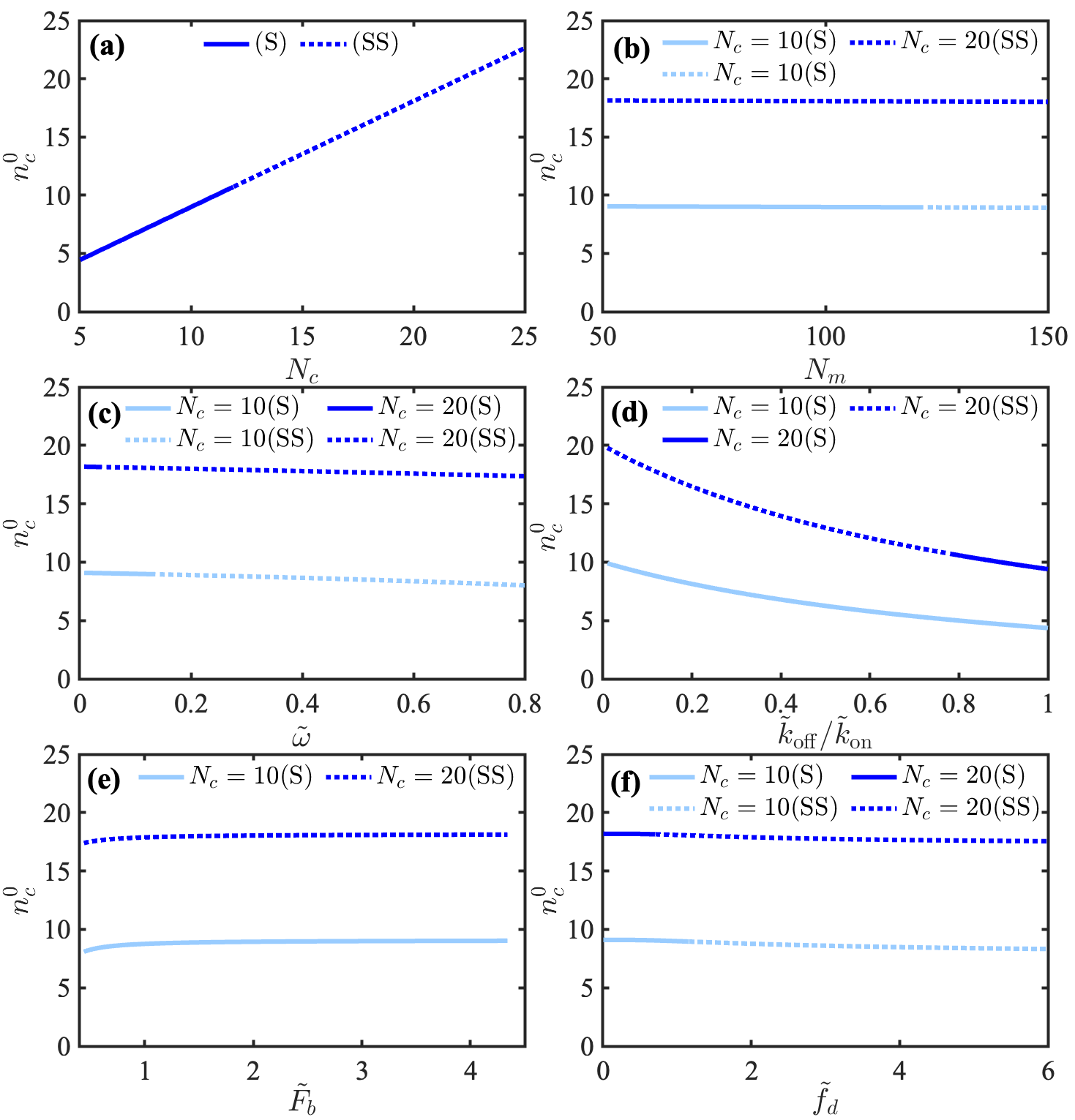}\\
	\caption{\label{nc} 
The impact of parameters variations on $n_{c}^{0}$ within the S and SS regions. Light-colored lines represent the scenario at $N_{c}=10$, while dark-colored lines are used for $N_{c}=20$. Solid and dashed lines denote their nature as stable (S) and stable spirals (SS), respectively.}
\end{figure}

The expression for $\tilde{x}_o^0$ includes $n_c^0$ and $n_{m}^0$, which in turn are related to multiple parameters, making it difficult to discern the individual effects of each parameter on $\tilde{x}_o^0$.    
To clarify these relationships, Fig.~\ref{xo} presents trend graphs illustrating the influence of parameters.
In Fig.~\ref{xo}(a), we can see that $\tilde{x}_o^0$ monotonically decreases with an increase in $N_c$.     
The system resides in the S state at lower $N_c$ values but transitions to the SS state upon exceeding a threshold of 11.92.
In Fig.~\ref{xo}(b), $\tilde{x}_o^0$ monotonically increases with $N_m$.
Fig.~\ref{xo}(c) shows that $\tilde{x}_o^0$ monotonically increases with $\tilde{\omega}$.
Notably, as $\tilde{\omega}$ approaches $0.8(\tilde{\omega}>0.76)$, we observe that the lighter line is shorter than the darker line, indicating that the system enters the unstable US state earlier.
This implies that an increase in $N_c$ shortens the region of state S and lengthens that of SS, and as $\tilde{\omega}$ increases, systems with a smaller $N_c$ more readily enter the unstable US state.

Fig.~\ref{xo}(d) illustrates that $\tilde{x}_o^0$ increases monotonically with $\tilde{k}_{\rm off } / \tilde{k}_{\rm on}$.  Specifically, for $N_c=10$, the system remains in the S state across the [0,1] range of $\tilde{k}_{\rm off } / \tilde{k}_{\rm on}$.  For $N_c=20$, the S state is observed for $0.78<\tilde{k}_{\rm off} / \tilde{k}_{\rm on}<1$, whereas the SS state is noted for $0<\tilde{k}_{\rm off} / \tilde{k}_{\rm on}<0.78$.
In Fig.~\ref{xo}(e), $\tilde{x}_o^0$ monotonically decreases as $\tilde{F}_b$ increases. With $N_c=10$, the fixed point does not exist for $\tilde{F}_b < 0.27$. In the range of $\tilde{F}_b$ from 0.27 to 4.34, the system consistently remains in the S state. For $N_c=20$, the fixed point disappears when $\tilde{F}_b < 0.13$, and the system enters the SS state within the range [0.13, 4.34].
Lastly, in Fig.~\ref{xo}(f), $\tilde{x}_o^0$ monotonically increases with $\tilde{f}_d$.

\begin{figure}[htbp]
	\includegraphics[scale=0.34]{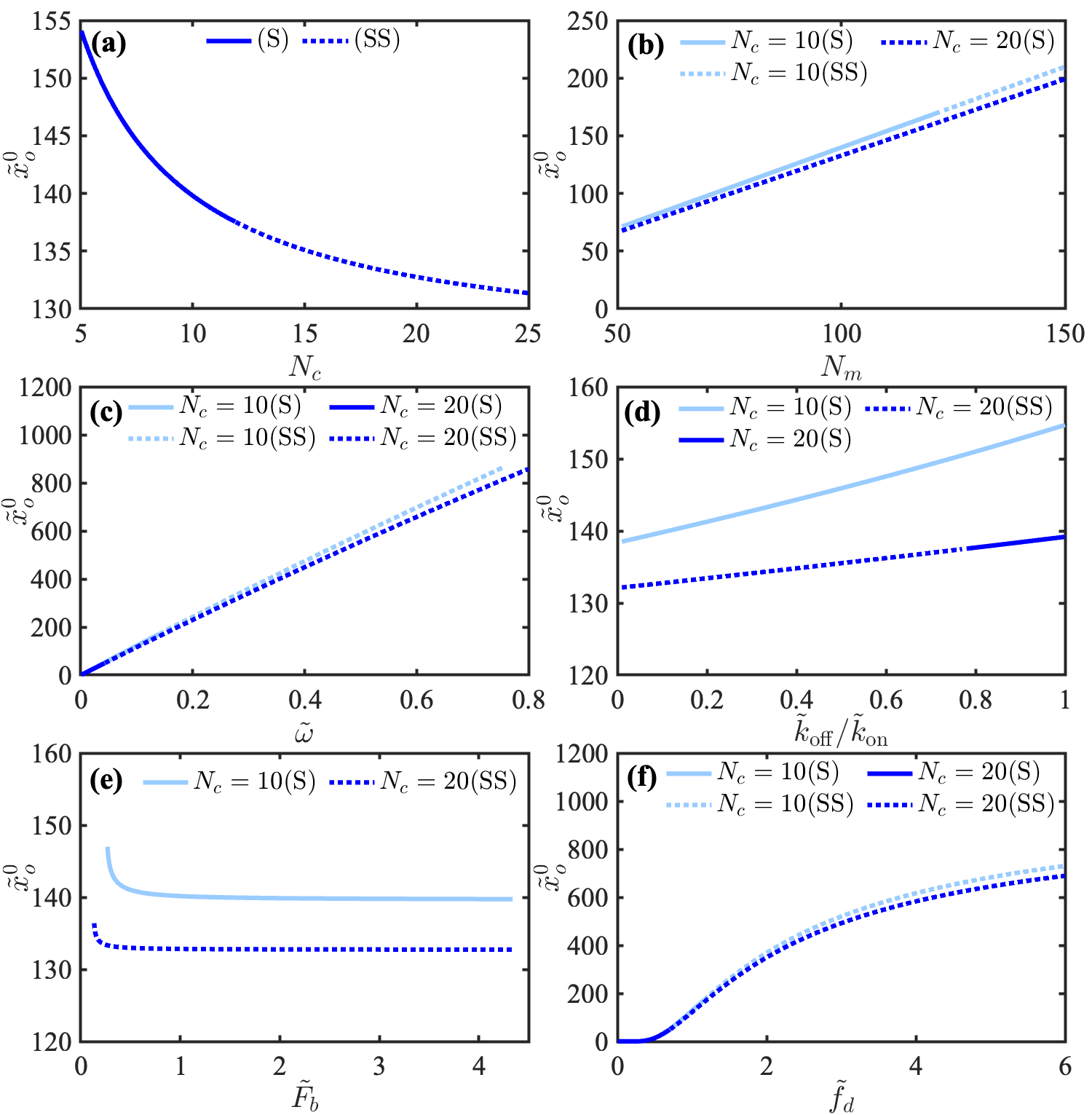}\\
	\caption{\label{xo} The impact of variations in parameters on  $\tilde{x}_{o}^{0}$ within the S and SS regions.
	Light-colored lines denote the case at $N_{c}=10$, whereas dark-colored lines correspond to $N_{c}=20$. Solid lines indicate a stable (S) nature, while dashed lines represent stable spirals (SS).}
\end{figure}

Since the situation for $\tilde{x}_c^0$ is analogous to that of $\tilde{x}_o^0$, the graphs for $\tilde{x}_c^0$ are omitted here.
Unlike the monotonic influence observed with other parameters, $\tilde{f}_s$ exhibits a non-monotonic effect on the fixed points. Consequently, we specifically illustrate this unique impact through a graph, as demonstrated in Fig.~\ref{fs}.

\subsection{The Analysis of the Unstable Spirals Region}
As previously mentioned, within the unstable spiral region, there exists a stable limit cycle around an unstable fixed point, corresponding to the system's periodic oscillations. 
Here, we investigate how the frequency and amplitude of the limit cycle vary with the parameter $\tilde{v}_u$ within the unstable spiral region in Fig.~\ref{Nc_vu}(a).

\begin{figure}[htbp]
	\includegraphics[scale=0.38]{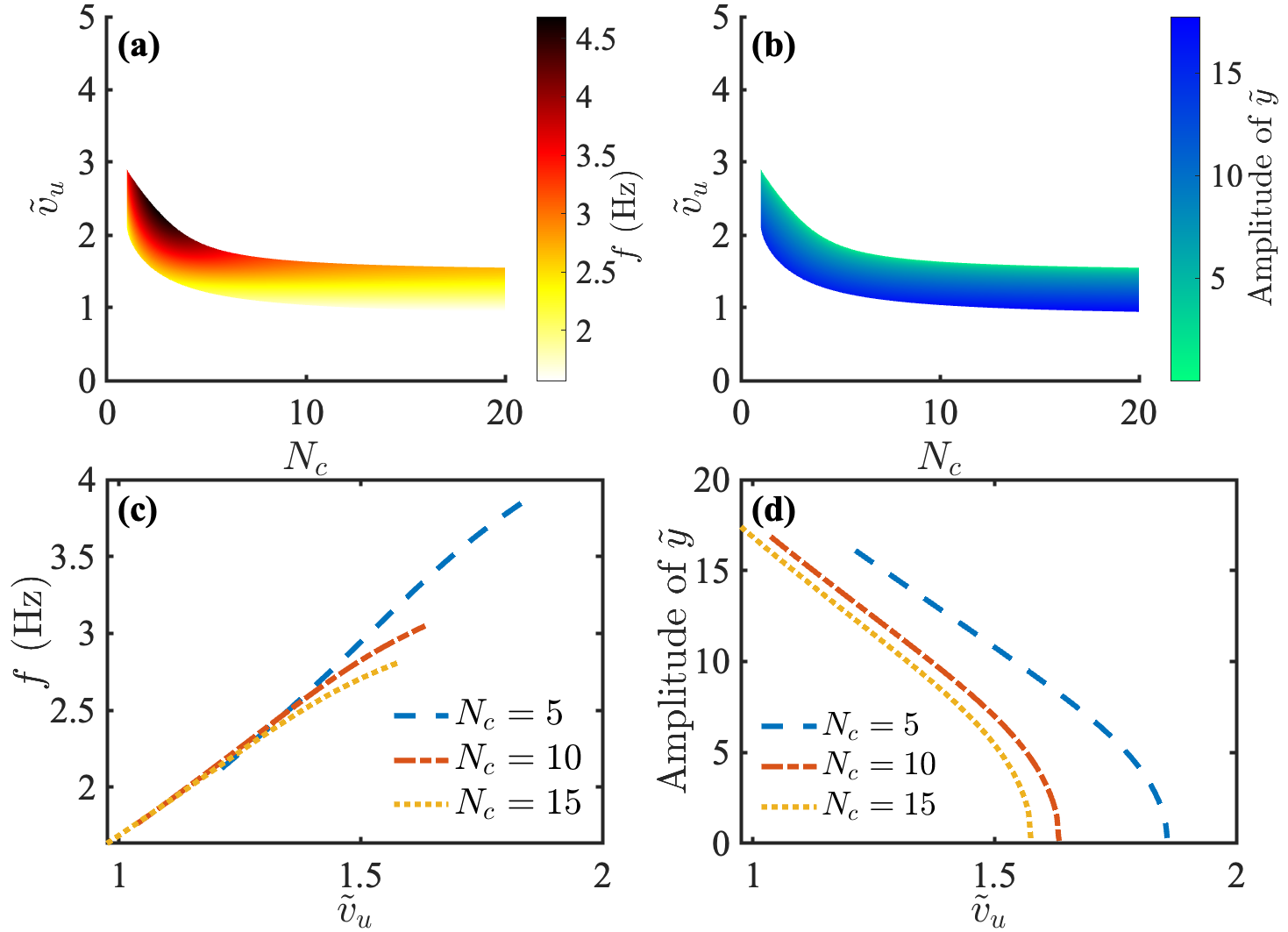}\\
	\caption{\label{Fre_yuntu}
		Frequency and amplitude of the limit cycle in the unstable spirals region.
		We focus on the unstable spirals region (US) from Fig.~\ref{Nc_vu}(a).
		To enhance visibility, we narrow the range of the $\tilde{v}_{u}$ axis from 0-15 to 0-5.	
		(a) Heat map showing the variation of frequency $f $ with changes in $\tilde{v}_{u}$ and $N_{c}$, with the color gradient representing the frequency in hertz (Hz).
		(b) Heat map depicting the variation of amplitude of $\tilde{y}$ with $\tilde{v}_{u}$ and $N_{c}$, with the color gradient representing the magnitude of amplitude. 
		(c) Variation of frequency $f $ with $\tilde{v}_{u}$ at fixed values of $N_{c}=$5, 10, and 15.
		(d) Variation of amplitude of $\tilde{y}$ with $\tilde{v}_{u}$ at fixed values of $N_{c}=$5, 10, and 15.
		(c) and (d) provide detailed line profiles extracted from the heat maps in (a) and (b), offering a more intuitive analysis at specific. To see this figure in color, go online.}
\end{figure}

As shown in Fig.~\ref{Fre_yuntu}, the frequency and amplitude of the limit cycles within the unstable spirals region are determined through numerical calculations.
As depicted in Fig.~\ref{Fre_yuntu}(a) and (c), the oscillation  frequency $f$ increases with an increase in velocity 
 $\tilde{v}_u$, while it decreases as $N_c$ increases.

In Fig.~\ref{Nc_vu}(a), on the upper boundary of the US region, where the real part of the complex conjugate roots $\alpha=0$, the corresponding $\tilde{v}_u$ is a function of $N_c$.
This critical value of $\tilde{v}_{u}$ is denoted as $\tilde{v}_{u}^{0}\left(N_c\right)$.
When $N_c$ is fixed, if $\tilde{v}_u<\tilde{v}_u^0\left(N_c\right)$, then as $\tilde{v}_u$ decreases, the amplitude of $\tilde{y}$ is proportional to $\sqrt{\tilde{v}_u^0\left(N_c\right)-\tilde{v}_u}$, as shown in Fig.~\ref{Fre_yuntu}(d).
As $\tilde{v}_u$ decreases, the amplitude of $\tilde{y}$ increases, engendering increasingly pronounced oscillations and augmented system instability, ultimately causing the system to enter the unstable region, as shown in Fig.~\ref{Fre_yuntu}(b). Simultaneously, the decrease in frequency $f$ accompanies the decrease in $\tilde{v}_u$, extending the oscillatory periods.

\subsection{The Effect of External Load on Dynamical Systems}

\begin{figure}[htbp]
	\includegraphics[scale=0.38]{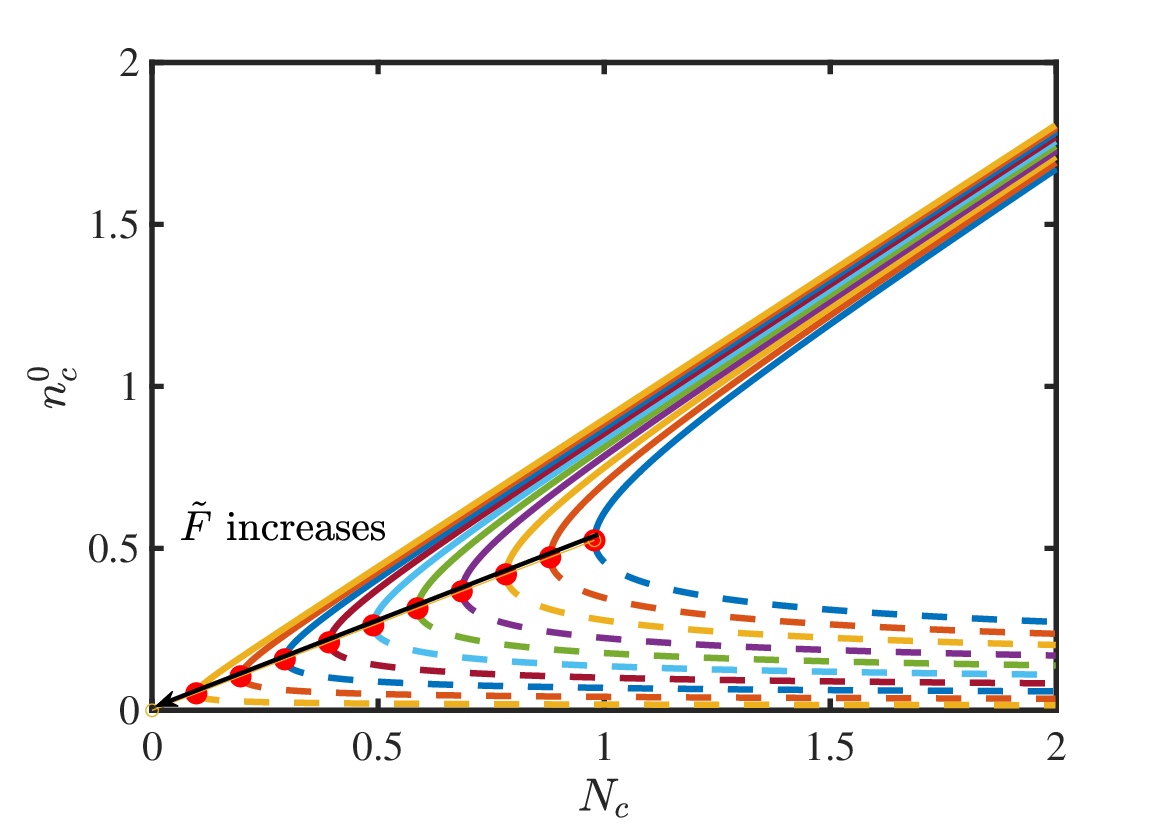}\\
	\caption{\label{Nc_nc_F}The influence of external load $\tilde{F}$ on saddle-node Bifurcation. 
	The direction of the arrow in the figure represents  the increasing direction of the external load $\tilde{F}$, where the starting point of this arrow corresponds to an external load $\tilde{F}=0$. 
	Following the direction of the arrow, each point corresponds to an external load $\tilde{F}=\frac{j}{10}n_m^0\tilde{f}_s$, where $j$ ranges from 0 to 9.}
\end{figure}

 Based on Fig.~\ref{Nc_nc_F}, it is observed that the critical value 
 $N_{c, {\rm min}}$ is dependent on the external load $\tilde{F}$.
 Within the interval $0 \leq \tilde{F}<n_m^0 \tilde{f}_s$, we note that $N_{c, {\rm min}}$ monotonically decreases to 0 as $\tilde{F}$ increases. This figure incorporates ten distinct lines which, following the direction of the arrows, correspond to gradually increasing external loads $\tilde{F}=\frac{j}{10}n_m^{0}\tilde{f}_s$, where $j$ ranges from 0 to 9.
 The starting point of this arrow corresponds to the external load of $\tilde{F}=0$, which is equivalent to the line in Fig.~\ref{nc_Nc}. This facilitates a comprehensive visual representation of how $N_{c, {\rm min}}$ is influenced by varying degrees of external load. 
 
\begin{figure}[htbp]
	\includegraphics[scale=0.38]{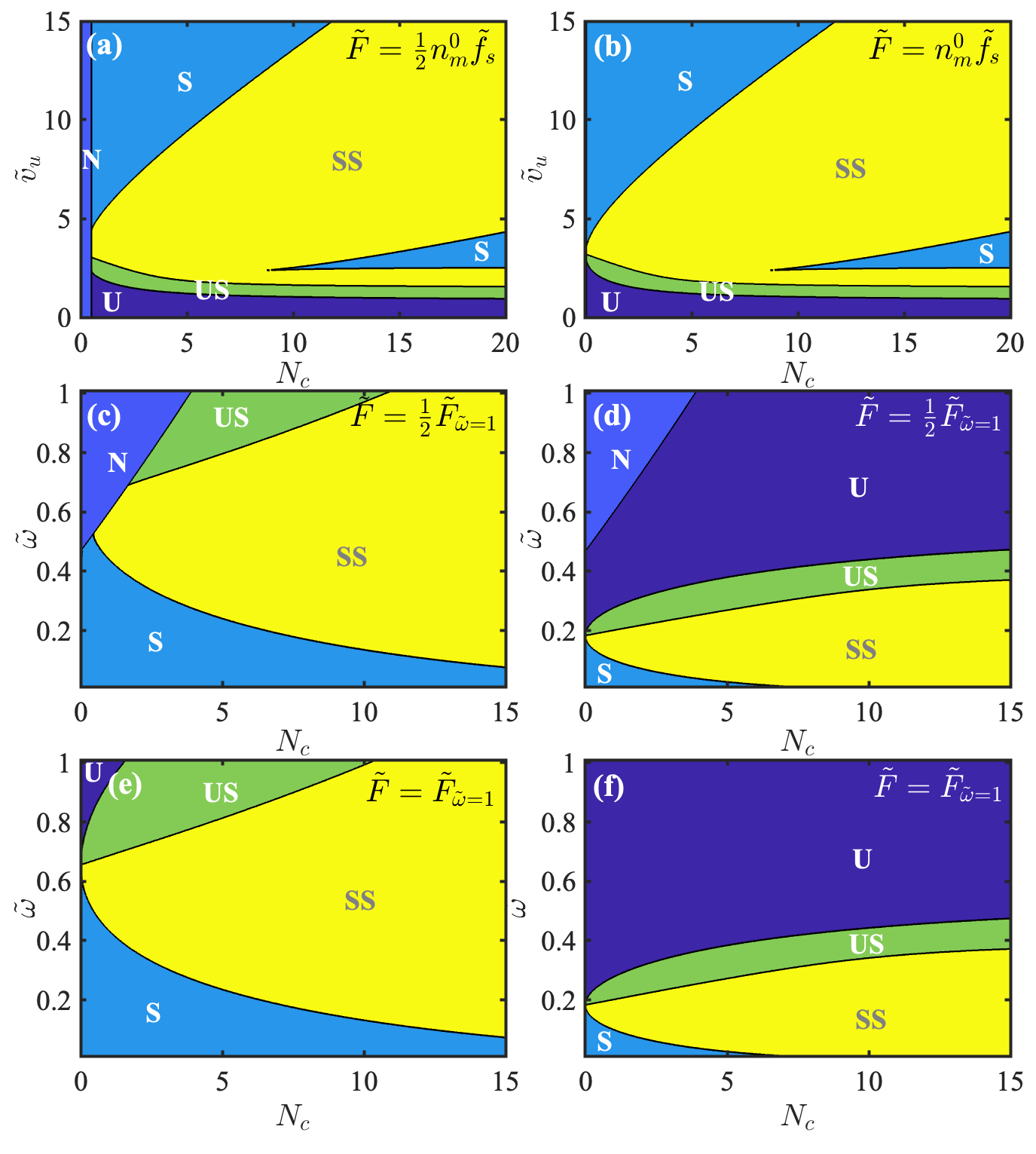}\\
	\caption{\label{Nc_F}The influence of external load $\tilde{F}$ on system stability.
(a) and (b) represent the stability diagrams at $\tilde{F}=\frac{1}{2}n_m^{0}\tilde{f}_s$ and $\tilde{F}=n_m^{0}\tilde{f}_s$, respectively, for Fig.~\ref{Nc_vu}(a).
		(c) and (d) illustrate the stability diagrams for Fig.~\ref{omega_Nc} under an external load $\tilde{F}=\frac{1}{2} \tilde{F}_{\tilde{w}=1}$.
		(e) and (f) depict the stability diagrams for Fig.~\ref{omega_Nc} with an external load $\tilde{F}=\tilde{F}_{\tilde{w}=1}$.}
\end{figure}

According to Figs.~\ref{Nc_F}(a) and (b), we observe that as the external load $\tilde{F}$ increases, the region N, where the fixed point does not exist, gradually diminishes until it disappears. 
This is equivalent to the original image being translated to the left. This phenomenon can be elucidated by referencing the balance equation $n_m\tilde{k}_m\tilde{y}=n_c\tilde{k}_c\tilde{x}_c+\tilde{F}$. As $\tilde{F}$ increases, it takes over a part of the force generated by myosin, leading to a reduction in the force borne by the clutch; that is, $\tilde{x}_c$ becomes smaller. Consequently, the detachment rate $\tilde{k}_{\rm off}$ decreases. 
According to the equation 
$0=\tilde{k}_{\rm on }\left(N_c-n_c\right)-\tilde{k}_{\rm off}^0 n_c \exp \left(\frac{\tilde{\kappa}_c\tilde{x}_c}{\tilde{F}_b}\right)$, the corresponding critical value $N_{c, {\rm min}}$ also decreases, hence the region N becomes smaller. The diminution of the N region as $\tilde{F}$ increases reflects the system's adaptive response to external mechanical stimuli.

Given that $n_m^{0}$ varies with changes in $\tilde{\omega}$, resulting in a variable $\tilde{F}=n_m^{0}\tilde{f}_s$, a constant external load force is considered by setting $\tilde{\omega}=1$. 
At this condition, the corresponding $$\tilde{F}_{\tilde{w}=1}=\frac{N_m}{1+\exp \left(\tilde{f}_s / \tilde{f}_d\right)} \tilde{f}_s,$$ thus facilitating a standardized analysis with external loads set at $\tilde{F}=\frac{1}{2} \tilde{F}_{\tilde{w}=1}$ and $\tilde{F}=\tilde{F}_{\tilde{w}=1}$ respectively.

Figs.~\ref{Nc_nc_F}(c), (d), and (e), (f) correspond to the scenarios of Fig.~\ref{omega_Nc} under external loads $\tilde{F}=\frac{1}{2} \tilde{F}_{\tilde{w}=1}$ and $\tilde{F}=\tilde{F}_{\tilde{w}=1}$, respectively. Similarly, we observe the region N contracts and ultimately vanishes as the external load $\tilde{F}$ increases, due to the critical value $\tilde{\omega}_{\rm max}(N_c, \tilde{F})$ increasing with $\tilde{F}$. As previously mentioned, according to the balance equation, the increase in $\tilde{F}$ leads to a diminution of the force endured by the clutch, resulting in a decrease in $\tilde{x}_c$. Consequently, the detachment rate $\tilde{k}_{\rm off }$ decreases, and $n_c$ increases. This indicates that an increase in $\tilde{F}$ exerts a facilitative influence on clutch attachment, promoting the adhesion of a greater number of clutches.
Now, with $\tilde{F}$ offsetting a portion of this force, it implies that a greater force generated by myosin (i.e., an increase in $n_m$) is needed to equate the force experienced by the clutches to it original level, thereby achieving the critical state value, corresponding to an increase in $\tilde{\omega}_{\rm max}(N_c, \tilde{F})$. This phenomenon is manifested as the diminution of the N region.

An increase in the external load $\tilde{F}$ effectively substitutes for the clutches by sharing a portion of the force produced by myosin motors, resulting in a stabilizing effect and leading to an expansion of the system's region of stability. This occurs because the increase in force from myosin induces an increase in $\tilde{x}_c$ and consequently elevates $\tilde{k}_{\rm off}$, leading to an expedited detachment of the clutches. This is identified as the underlying cause for the disappearance of fixed points. The  increase in $\tilde{F}$ serves to mitigate this destabilizing influence.

Furthermore, according to the fixed-point expressions Eq.~\eqref{y0}-\eqref{xc0}, we know that an increase in $\tilde{F}$ affects the position of the fixed points.
To summarize, we find that an  increase in the external load $\tilde{F}$ primarily results in the reduction of the N region, which is equivalent to a translation of the original image.

\section{Conclusion}

In this work, we introduce a five-dimensional nonlinear autonomous system for describing and simulating the dynamics of intracellular actomyosin activity, as well as the complex biomechanical interactions between the cell and its external environment. This system model is designed to investigate the influences of myosin, clutches, substrate, and external load on the system's stability.
 Moreover, we explore the oscillatory behavior within the unstable spiral region and analyze the effects of various parameters on fixed points within both the stable and stable spiral regions.

As predicted by previous models, myosin contractility is a key candidate for causing fluctuations due to its ability to induce spontaneous oscillations \cite{grill2005theory}.  Moreover, there is experimental evidence of myosin's role in the specific context of force fluctuations within individual focal adhesions (FAs) and in the regulation of migration and mechanosensing \cite{wu2017two,greenberg2016perspective,pasapera2015rac1}.
Through the proposed system, which is applicable across a wide range of experimentally relevant parameters, we validate the significant influence of myosin activity on oscillatory behaviors.
We recognize that the principal factors which considerably influence the force fluctuations in FAs are myosin active velocity $\tilde{v}_u$ and the ratio $\tilde{w}$ of motor attachment rate to motor detachment rate. 
Our model reproduces stick-slip type behavior at lower active velocities $\tilde{v}_u$ or when the ratio $\tilde{w}$ is relatively high, and it successfully demonstrates self-sustaining oscillations known to occur within  (FAs)
 \cite{plotnikov2012force,wu2017two}.
In this study, we direct our attention to the unstable spirals region within the $N_c-\tilde{v}_u$ plane, exploring the oscillation  frequency and amplitude of the limit cycle. 
When $\tilde{v}_u$ falls below $\tilde{v}_u^0\left(N_c\right)$, a reduction in $\tilde{v}_u$ results in an increase in the amplitude of $\tilde{y}$, which is directly proportional to $\sqrt{\tilde{v}_u^0\left(N_c\right)-\tilde{v}_u}$, thus giving rise to more pronounced oscillatory behaviors.
Concurrently, a decrease in the oscillation frequency $f$ is observed, leading to prolonged oscillatory periods.

Within our study, the N region emerges under two distinct conditions.
The first condition arises when $N_c$ falls below a minimal threshold, specifically $N_c<N_{c, \min }$. 
The second condition arises when $\tilde{\omega}$ exceeds its maximum allowable value, satisfying $\tilde{\omega}>\tilde{\omega}_{\rm max}(N_c, \tilde{F})$.

We discuss the stability of the dynamic system from four aspects: myosin motors, molecular clutches, compliant substrate, and external load.
When solution exist, our findings indicate that an increase in the velocity of $\tilde{v}_u$ tends to enhance the system's stability.
A higher ratio of $\tilde{\omega}$ leads to an increased number of attached motors $n_m$, thereby augmenting the force exerted by myosin and rendering the system more unstable.    Furthermore, the lower the stiffness of the clutch, the more unstable the system becomes.

Both substrate stiffness and substrate friction coefficient influence the system similarly by hindering the movement of substrate linkages.
The smaller the substrate stiffness and substrate friction coefficient, the lesser the hindrance, making the system more unstable;     conversely, the greater the hindrance, the more easily the system stabilizes.
Thus, augmenting substrate stiffness and substrate friction coefficient plays a stabilizing role in the system's dynamics.

The external load influences the position of fixed points.
The increase in $\tilde{F}$ alleviates a fraction of the force exerted by myosin motors, thereby diminishing the force endured by the clutches.
Within a certain range, an increase in $\tilde{F}$ results in a diminishment and eventual disappearance of the N region, consequently expanding the area of regions where solutions exist.

The presence of an external load or a substrate could change the timescales of attachment/detachment of myosin motors and clutch proteins \cite{ghosh2022deconstructing}. 
This means that the existence of a substrate or the application of an external load could accelerate or decelerate the dynamic interactions between proteins, thereby influencing the cellular mechanisms of sensing and responding to the external environment.
An external load applied to actin filaments affects the interaction between myosin and clutch proteins. 
The application of the external load slows down the detachment process of the clutch, thereby facilitating an increased adherence of clutch proteins.

\section*{ACKNOWLEDGMENTS}
This research is supported by the Science and Technology Commission of Shanghai Municipality (23JC1400501) and  the Natural Science Foundation of China (12241103).

%\hrulefill\dotfill\rule{20em}{0.2em}\dotfill\hrulefill
%\rule{26em}{0.2em}
%
%
%\hrulefill\dotfill\rule{20em}{0.2em}\dotfill\hrulefill
%\rule{26em}{0.2em}

%\clearpage
%\newpage
%\pagebreak
%\vspace*{1cm}

%\begin{appendix}
\appendix
\setcounter{figure}{0}
\setcounter{section}{0}
\setcounter{equation}{0}
\setcounter{table}{0}
\renewcommand{\thefigure}{A\arabic{figure}}
\renewcommand{\thesection}{A\arabic{section}}
\renewcommand{\theequation}{A\arabic{equation}}
\renewcommand{\thetable}{A\Roman{table}}

\begin{widetext}
\maketitle
\section*{\bf APPENDIX: COMPUTING THE JACOBIAN MATRIX ABOUT THE FIXED POINTS OF THE SYSTEM}\label{appendex}
\vspace*{0.3cm}

In this Appendix, we rewrite five-dimensional nonlinear system Eqs.~\eqref{Dimensionless_Eq} in vector form,

\begin{equation}
	\frac{d}{d \tau}\left(\begin{array}{c}
		n_m \\
		n_c\\
		\tilde{x}_o \\
		\tilde{y} \\
		\tilde{x}_s
	\end{array}\right)=\left(\begin{array}{c}
		\tilde{\omega}\left(N_m-n_m\right)-n_m \exp \left(\frac{\tilde{\kappa}_m \tilde{y}}{\tilde{f}_d}\right)\\
		\tilde{k}_{\text {on }}\left(N_c-n_c\right)-\tilde{k}_{\text {off }} n_c \exp \left(\frac{\tilde{\kappa}_c \left( \tilde{x}_o-\tilde{x}_s \right) }{\tilde{F}_b}\right)\\
		n_m \tilde{\kappa}_m \tilde{y}-n_c \tilde{\kappa}_c \tilde{x}_c-\tilde{F} \\
		\tilde{v}_u\left(1-\frac{\tilde{\kappa}_m \tilde{y}}{\tilde{f}_s}\right)-\left(n_m \tilde{\kappa}_m \tilde{y}-n_c \tilde{\kappa}_c \tilde{x}_c-\tilde{F} \right) \\
		\frac{\Gamma}{\gamma}\left[ n_c \tilde{\kappa_c} \left( \tilde{x}_o-\tilde{x}_s\right)-k_s \tilde{x}_s\right] 
	\end{array}\right),
	\label{vector}
\end{equation}
we denote the vector on the right-hand side of Eqs.~\eqref{vector} as the  vector {\bf f}, where its components are $f_1, f_2, \cdots, f_5$.

To analyze the stability near fixed points, we perform a linear approximation of this five-dimensional nonlinear system.
Below is the linear system we introduce. To distinguish it from the preceding nonlinear system, we rename the variables as $\delta n_m, \delta n_c, \delta \tilde{x}_o, \delta \tilde{y}, \delta \tilde{x}_s$,
\begin{equation}
	\frac{d}{d \tau}\left(\begin{array}{c}
		\delta n_m \\
		\delta n_c\\
		\delta \tilde{x}_o \\
		\delta \tilde{y} \\
		\delta \tilde{x}_s
	\end{array}\right)=\mathcal{J}\left(\begin{array}{c}
		\delta n_m \\
		\delta n_c\\
		\delta \tilde{x}_o \\
		\delta \tilde{y} \\
		\delta \tilde{x}_s
	\end{array}\right)=\left[\begin{array}{ccccc}
	\frac{\partial f_1}{\partial n_m} &\frac{\partial f_1}{\partial n_c} &\frac{\partial f_1}{\partial \tilde{x}_o} &	\frac{\partial f_1}{\partial \tilde{y}} & \frac{\partial f_1}{\partial \tilde{x}_s} \\
	\frac{\partial f_2}{\partial n_m} &\frac{\partial f_2}{\partial n_c} &\frac{\partial f_2}{\partial \tilde{x}_o} &	\frac{\partial f_2}{\partial \tilde{y}} & \frac{\partial f_2}{\partial \tilde{x}_s} \\
	\frac{\partial f_3}{\partial n_m} &\frac{\partial f_3}{\partial n_c} &\frac{\partial f_3}{\partial \tilde{x}_o} &	\frac{\partial f_3}{\partial \tilde{y}} & \frac{\partial f_3}{\partial \tilde{x}_s} \\
	\frac{\partial f_4}{\partial n_m} &\frac{\partial f_4}{\partial n_c} &\frac{\partial f_4}{\partial \tilde{x}_o} &	\frac{\partial f_4}{\partial \tilde{y}} & \frac{\partial f_4}{\partial \tilde{x}_s} \\
	\frac{\partial f_5}{\partial n_m} &\frac{\partial f_5}{\partial n_c} &\frac{\partial f_5}{\partial \tilde{x}_o} &	\frac{\partial f_5}{\partial \tilde{y}} & \frac{\partial f_5}{\partial \tilde{x}_s} 
\end{array}\right] \left(\begin{array}{c}
\delta n_m \\
\delta n_c\\
\delta \tilde{x}_o \\
\delta \tilde{y} \\
\delta \tilde{x}_s
\end{array}\right).
\end{equation}

We compute the Jacobian matrix about the fixed points of the system,
  \begin{equation}
 	\mathcal{J}
 	=\left[\begin{array}{ccccc}
 		-\tilde{\omega}-\exp \left(\frac{\tilde{f}_s}{\tilde{f}_d}\right)&0&0&-n_m^0 \frac{\tilde{\kappa}_m}{\tilde{f}_d} \exp \left(\frac{\tilde{f}_s}{\tilde{f}_d}\right) &0\\
 		0&-\tilde{k}_{\rm on }-\tilde{k}_{\rm off} \exp \left(\frac{n_m^0 \tilde{f}_s-\tilde{F}}{n_c^0 \tilde{F}_b}\right) &-\tilde{k}_{\rm off }\frac{n_c^0 \tilde{\kappa}_c}{\tilde{F}_b} \exp \left(\frac{n_m^0 \tilde{f}_s-\tilde{F}}{n_c^0 \tilde{F}_b}\right)&0&\tilde{k}_{\rm off }\frac{n_c^0 \tilde{\kappa}_c}{\tilde{F}_b} \exp \left(\frac{n_m^0 \tilde{f}_s-\tilde{F}}{n_c^0 \tilde{F}_b}\right)\\
 		\tilde{f}_s& -\frac{n_m^0 \tilde{f}_s-\tilde{F}}{n_c^0} &  -n_c^0 \tilde{\kappa}_c  &  n_m^0 \tilde{\kappa}_m  &n_c^0 \tilde{\kappa}_c\\
 		-\tilde{f}_s &\frac{n_m^0 \tilde{f}_s-\tilde{F}}{n_c^0 }&  n_c^0 \tilde{\kappa}_c &  -\tilde{v}_u \frac{\tilde{\kappa}_m}{\tilde{f}_s}-n_m^0 \tilde{\kappa}_m  &-n_c^0 \tilde{\kappa}_c\\
 		0 &\frac{\Gamma}{\gamma}\frac{n_m^0 \tilde{f}_s-\tilde{F}}{n_c^0}&\frac{\Gamma}{\gamma}n_c^0 \tilde{\kappa}_c& 0&-\frac{\Gamma}{\gamma}n_c^0 \tilde{\kappa}_c- \tilde{\kappa}_s
 	\end{array}\right].
 \end{equation}

\begin{figure}[htbp]
	\includegraphics[scale=0.38]{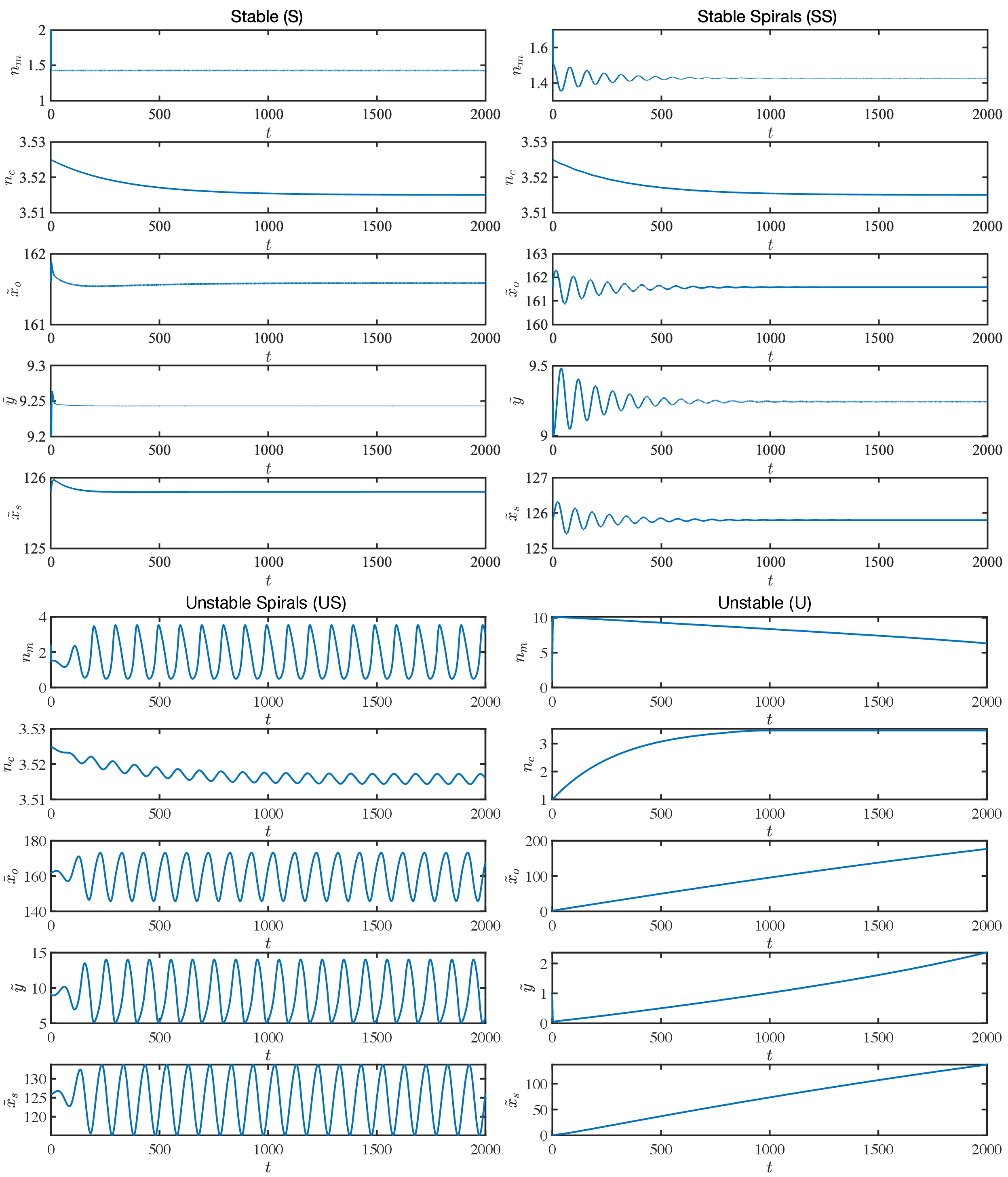}\\
	\caption{\label{validate}	Four dynamical behaviors of the system as identified in Tab.~\ref{tab:1}.
		In the different stability regions of the stability diagram on the $N_c-\tilde{v}_u$ plane (Fig.~\ref{Nc_vu}(a)), we selected a point from each region and plotted the numerical solutions of the dynamical equations at these points to validate our predictions made by linear stability analysis.}
\end{figure}

\begin{figure}[htbp]
	\includegraphics[scale=0.5]{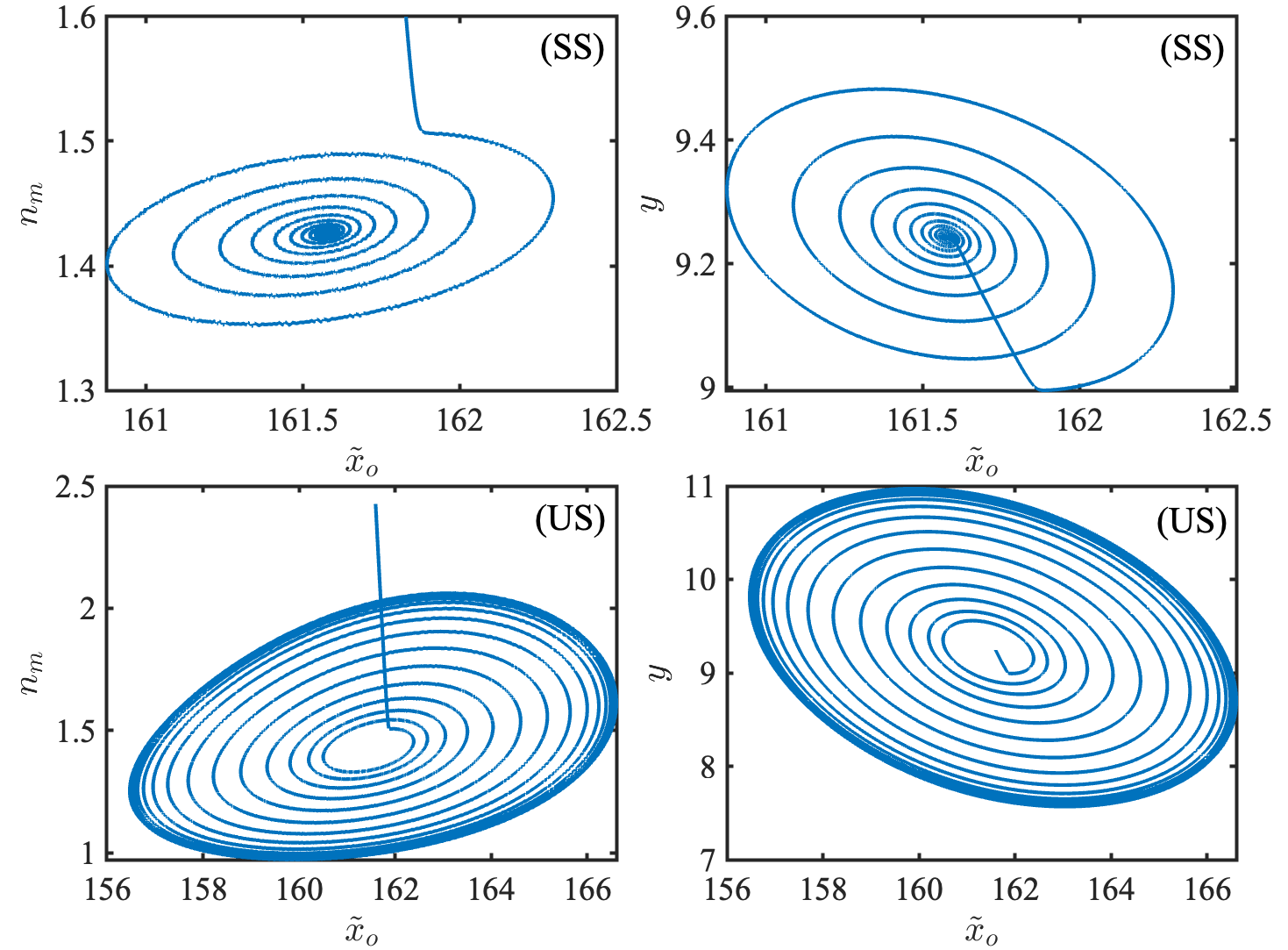}\\
	\caption{\label{Trajectory} Trajectory diagrams in the $n_{m}-\tilde{x}_{o}$ plane and $\tilde{y}-\tilde{x}_{o}$ plane under SS and US states.	}
\end{figure}

\begin{figure}[htbp]
	\includegraphics[scale=0.34]{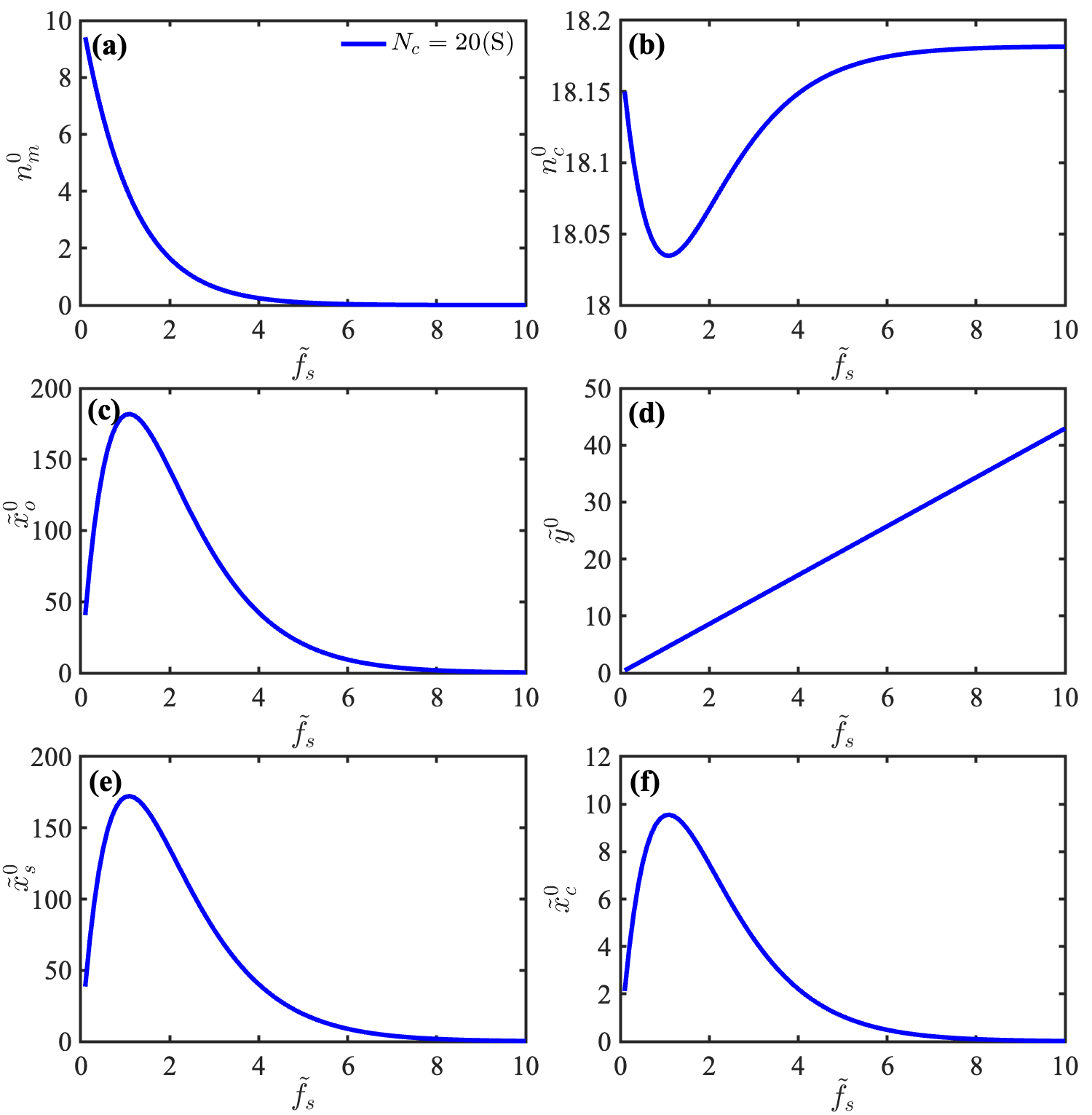}\\
	\caption{\label{fs} The effect of variations in the parameter $\tilde{f}_{s}$
		on the system's fixed point within the stable region at $N_{c}=20$.}
\end{figure}

%\bibliographystyle{unsrt}
%\bibliography{mybibtex}

\end{widetext}
\clearpage
\newpage
\bibliographystyle{unsrt}
\bibliography{mybibtex}
\end{document}